\documentclass[epj]{svjour}
\usepackage{graphics}
\usepackage{geometry}                
\usepackage{graphicx}
\usepackage{amssymb}
\def\pppbarp{$pp/{\bar p}p$}
\def\qmax{q_{max}}
\def\shat{${\hat s}$}
\def\ptmin{p_{tmin}}
\def\p{$p$}
\def\ktmax{k_{tmax}}

\begin{document}
\title{Total photoproduction cross-section  at very high energy}
\author{R.M. Godbole\inst{1} \and A. Grau\inst{2} \and G. Pancheri\inst{3} \and Y.N. Srivastava\inst{4}
}                     
%
%
\institute{Centre for High Energy Physics, Indian Institute of Science,
 Bangalore, 560012, India. \and Departamento de F\'\i sica Te\'orica y del Cosmos and CAFPE, Universidad de Granada, 18071 Granada, Spain. \and INFN Frascati National Laboratories, Via Enrico Fermi 40, I-00044 Frascati, Italy. \and INFN and Physics Department, University of Perugia, Via A. Pascoli, I-06123 Perugia, Italy}
\date{Received: date / Revised version: date}
%
\abstract{
In this paper we apply  to  photoproduction  total cross-section  a model we have proposed for purely hadronic processes and  which is  based on    QCD mini-jets and soft gluon re-summation. We compare the predictions of our model with the HERA data
as well as with other models. For cosmic rays, our model predicts
substantially higher  cross-sections at TeV energies than models based
on factorization but lower than models  based on mini-jets alone, without soft gluons. We discuss the origin of this difference.
} 
\maketitle
\section{Introduction}
\label{intro}
  Cosmic ray experiments and planning for a future linear collider require knowledge of photon total cross-sections for values of  very high c.m. energy of the photon-proton or photon-photon system, in regions where no  data are available. In this paper we study the predictions of a mini-jet model, extending to the photon a model developed for proton-proton scattering \cite{lastPLB}. We  apply this model to photo-production, leaving to a later paper the application to photon-photon scattering. 

 Understanding the energy dependence of total hadro--nic
cross-sections continues to be an important issue in the study of strong
interactions per se \cite{martin}. Over the years, various descriptions of this energy
dependence have been given.
Some approaches have focused on how far
one can reach following the basic principles of analyticity, unitarity,
factorisation etc., without any recourse to the details of the particular
hadron involved, whereas at the other end of the spectrum, there are models
which include the fundamentals
 of QCD as far as possible and then
try to compute the cross-section in terms of measured properties of the
particular hadron. Of course, all descriptions have to be consistent with the
requirements of analyticity and unitarity. Most descriptions involve a few
``soft'' (non-perturbative) parameters, which can not be determined through
perturbative QCD. Again, basic symmetry, unitarity and factorisation arguments
may at times lead to certain relationships among these soft parameters for
various hadrons. Often they may be determined only through fits to the
experimental data and then one may only test approximate
relations among these indicated by general arguments. In short, understanding
the behaviour of the total hadronic cross-section and other soft quantities such
as multiplicities etc., from first principles, is an extremely challenging
problem and as stated before, one has different answers with varying degrees of
relationship to QCD.

Hadronic cross-sections for processes induced by the photon and
the hadronic structure of the photon itself, have played a very interesting
and important r\^ole,  in furthering the attempts to understand
the theoretical issues involved in the subject.
Photon-hadron interactions offer the theorists one more laboratory to test
their various ideas about computing ``soft''
quantities such as purely hadronic total cross-sections from basic principles.
Historically, it is the interaction of the highly virtual photon with the
hadron that offered the first glimpse of (almost free) quarks and later
provided basic evidence for perturbative QCD being the correct dynamics to
explain strong interactions in a certain kinematic domain.
 However, in the
present context, it is the photon structure function language
\cite{REVPHOTSTR}  used to
describe
interactions of the real or quasi--real photon (invariant mass square
$\sim 0$), with other hadrons or photon, that is of interest. In fact, the
structure function of a quasi real photon at large values of $x_\gamma$
and that of a highly virtual photon (with large values of $Q^2$ where
$- Q^2$ indicates the invariant mass square of the virtual photon) for
all values of $x_\gamma$, can be computed using perturbative QED and QCD
alone, for large values of momentum transfer square, $Q^2$ of the probe.
However, equally important is the (non perturbative) part of the real
(or quasi real) photon structure function at small $x_\gamma$ which is not
amenable
to perturbative QCD (PQCD) computations.

In this paper, we apply our eikonal mini-jet model augmented by soft gluon
re\textit{}summation, which has been successful in providing an acceptable description of
the $pp / p \bar p$ data, to the description of total cross-sections of photon
induced processes.
In  our model for the (purely hadronic) proton total cross-sections, we were
able to compute the relevant components in terms of basic QCD inputs such as
the experimentally measured parton densities and QCD subprocess cross-sections
along with a few non-perturbative parameters. Given the prior success, it
becomes of  interest
to see how the predictions  of our model, applied to the total hadronic
cross-sections of photon induced processes  and  using the experimentally
determined knowledge on the structure of the ``real'' photon, compare with
the data. We shall be mainly concerned with the issue of its energy dependence.

To recapitulate: in this paper we  explore  the effects of the hadronic structure
of the photon through studies of total cross-sections involving photons.
While at low energy, these cross-sections can be obtained  through
factorization and vector meson dominance, 
the high energy
range poses a different challenge.
We have argued in a number of papers
{\cite{REVPHOTSTR,rohinicorsetti,albert,lincoll}
that the  energy dependence of the photon induced processes  do  not 
follow from a straightforward application of factorization properties of
the total cross-sections. We shall discuss various factorization
results \cite{gribov,martinfacto,cudell,bsw,dl}
 and compare  some of them with   the HERA data \cite{h1,zeus}
as well as with predictions of   
our QCD eikonal model 
with  resummation, hereafter referred to as the BN model \cite{ourlast}.
The reasons for this nomenclature  
will be clear as we describe the model. Some of its details are summarized in two Appendices, so as not to overburden the reader with material published elsewhere.

\section{Total cross-sections: from pp to $\gamma \gamma$}

Experimentally, all total cross-sections rise asymptotically with
energy, but it is not yet clear whether the rate of increase is the same for
different  processes and whether their asymptotic behaviour
is already controlled
by
the Froissart-Martin \cite{froissart} bound. For any given total hadronic cross-section, this bound says that
asymptotically
\begin{equation}
\sigma_{tot} \le C (\log {s})^2.
\label{froissart}
\end{equation}

From a theoretical point of view, this bound is only valid for the scattering of hadrons. Attempts to extend it to  virtual photon process \cite{GLM} have resulted in a less restrictive bound, but which cannot be extended to real photons, leaving unanswered  the question of whether the bound of Eq. (\ref{froissart}) is valid also for  real photons.

Phenomenologically, the LEP data \cite{datagg} seem to indicate that the
slope with which  the total $\gamma\gamma$ cross-section rises is not the
same as in the proton case\cite{albert}.  This difference would spoil the
simplicity of the so-called Regge-Pomeron model, in which the high energy
rise is described through a single universal term \cite{dl}.
Of course, all total cross-sections do rise and to appreciate it at a glance,
we show in Fig. ~\ref{Fig:TX} a compilation of  data on \pppbarp\
\cite{PDG}\cite{dataproton}, $ \gamma p$ \cite{h1,zeus} and
$\gamma \gamma$ \cite{datagg}  scattering together with expectations from the BN model \cite{ourlast}  for protons  to be described in the next section. Since the data span an energy range of four
orders of magnitude, with the cross-sections in the millibarn range
for proton-proton, microbarn range for photoproduction and nanobarns for
photon-photon, to plot them all on the same scale, one needs a normalization
factor.  The data suggest to multiply the $\gamma p$ cross-section by a factor $\approx 330$ and then $\gamma \gamma $ by $(330)^2$, as shown in Fig. ~\ref{Fig:TX}.

 It has been known for quite some  time  \cite{collins} that  to get  the photoproduction cross-section from the proton cross-sections
in the region where they are approximately constant, namely after the initial Regge-exchange
type fall and before the beginning of the high energy rise, the multiplicative
factor to apply for each photon leg in the cross-section
can be obtained from Vector Meson Dominance (VMD) (to go from a photon to a meson) and a quark counting factor, namely
\begin{eqnarray}
R_{\gamma}&=&
 {{N^{fermion\ lines}_{meson}} \over {N^{fermion\ lines}_{proton}}} P_{VMD}={{2}\over{3}}
(\sum_{V=\rho,\omega,\phi} P_V ) \nonumber \\
&=& {{2}\over{3}} (\sum_{V=\rho,\omega,\phi}{{4\pi \alpha}\over{f_V^2}})
\label{rgamma}
\end{eqnarray}
 With present $\rho$-meson data \cite{PDG08} and
 the relation
\begin{equation}
P_{\rho} = {{e^2}\over{f^2_{\rho}}}={{\alpha}\over{12}} {{m_\rho}\over{\Gamma_{\rho}}}
\end{equation}
we would obtain  $R_{\gamma} \approx 1/360$, consistent with   the value  indicated in the figure.
%

\begin{figure}[htbp]
\centerline{\includegraphics*[width=1.0\columnwidth]{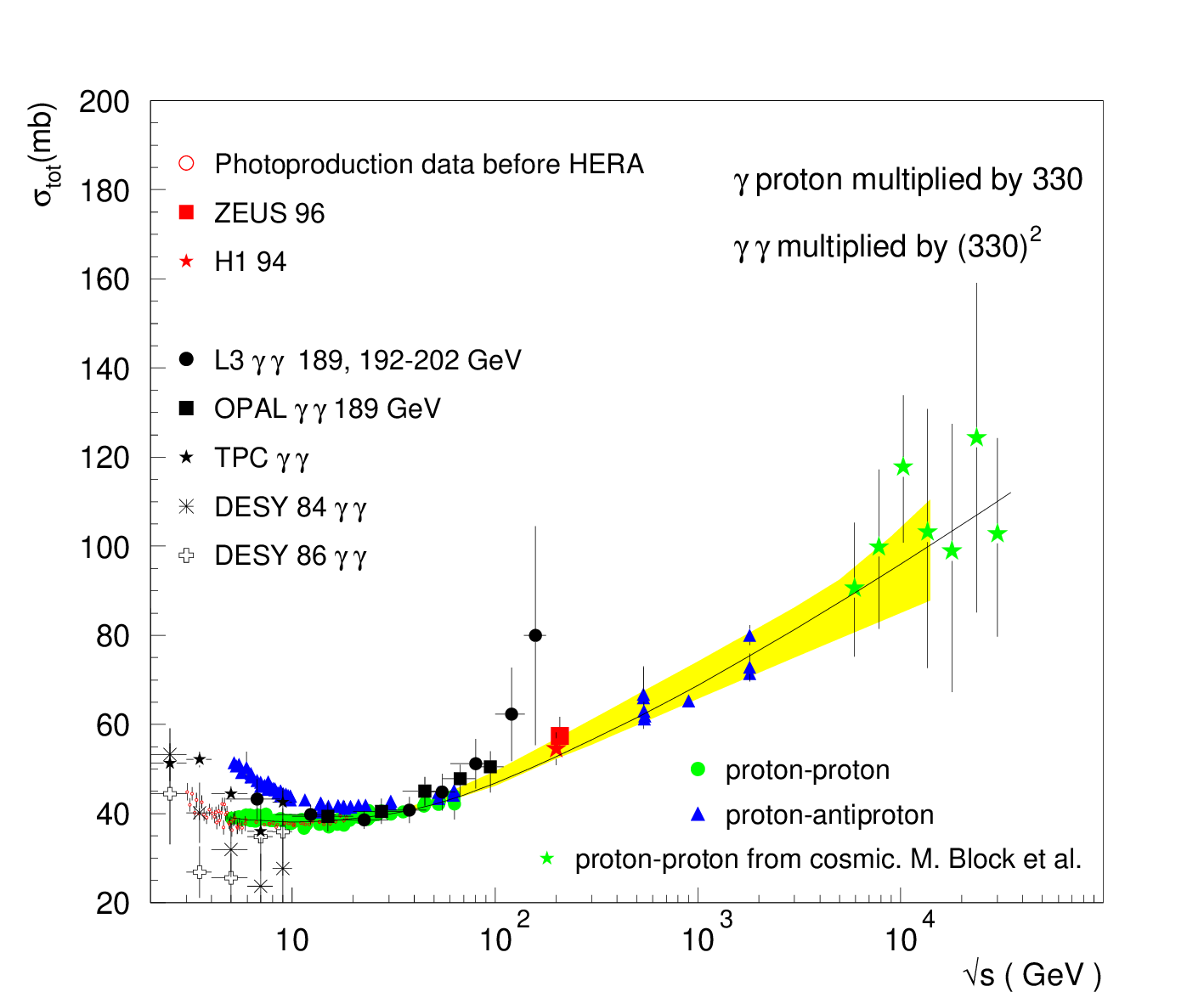}}
 \caption{Proton \cite{dataproton} and photon \cite{h1,zeus,datagg} normalized
total cross-sections with the band  expected from our BN model for \pppbarp  \  \cite{lastPLB} and one  typical curve from this model   \cite{ourlast}.  Cosmic ray data are from ref.~\cite{martincosmic}.}
\label{Fig:TX}
\end{figure}

 Note that
there is no a priori reason to expect the scaling factor to be energy
independent.
On the other hand, while at low energies the factor $R_{\gamma}$  can be reliably  evaluated  through VMD,
at high energies, it is likely to be different~\cite{desy}
due to the difference in the quark and gluon content of
photons~\cite{REVPHOTSTR} versus  that of the hadrons.
The use of a simple multiplicative factor to compare the photon processes with each other and with the pure proton processes, is the simplest form of {\it factorization}. More complex forms of factorization exist in the literature, as in a recent formulation by Vereshkov and collaborators \cite{vereshkov08} or as in  the model by Block et al. (also called the Aspen model) \cite{aspen}. More comments shall follow in the last section.

The above  discussion points to the need for a  description of high energy photon interactions
where reliable predictions can be made based on the  quark-parton structure of the photon.
As stated earlier, we have developed such a model for purely hadronic processes \cite{lastPLB,ourlast,our99,corsetti} which
we shall extend and apply to photoproduction processes in the next section.

\section{The Bloch-Nordsieck model (BN)}

This model is based on the eikonal representation for the total
cross-section \cite{chengwu,bswimpact}, and the eikonal incorporates QCD
inputs such as parton-parton cross-sections, 
parton densities  extracted from perturbative QCD fits to the data,
actual kinematics, and soft gluon resummation.  In detail, we use:
\begin{enumerate}
\item QCD mini-jets \cite{jacob,rubbia} to drive the rise of the total cross-section
in the QCD asymptotic freedom regime;
\item the eikonal representation for the total cross-section with the real part of the eikonal approximated to zero and the imaginary part obtained through mini-jet QCD cross-sections;
\item an   impact parameter distribution obtained as the  Fourier transform of the re-summed  soft gluon transverse momentum distribution;
\item resummation of soft gluon emission down to zero momentum  to
soften the rise due to the increasing number of gluon-gluon collisions
between low-x, but still hard perturbative, gluons.
\end{enumerate}
While the eikonal representation with the mini-jet input has long been in use, our model differs from other existing eikonal models in that the impact parameter distribution is energy dependent and derived from soft gluon $k_t$-resummation, which gives the model its name.

Before turning to the application of the model to photon processes, we shall briefly discuss the various approximations involved.
\begin{enumerate}
\item the energy dependence of QCD mini-jets reflects the increase -at low x- of the gluon structure functions and this can provide a mechanism for the rise of total cross-sections. However, parton-parton cross-sections  require a minimum transverse momentum cut-off to avoid the $1/p_t^2$ divergence. Fixing such a cut-off around 1 GeV,
leads to an extremely sharp rise. Our model starts with this input and then tempers the rise through the energy-dependent impact parameter distribution.
\item the eikonal representation for  total cross-sections is an approximation which
allows one to enforce the requirement of s-channel
unitarity.
Here one obtains the total cross-sections through the eikonal formulae
\begin{eqnarray}
\sigma_{tot}=2\int d^2{\vec b}[1-e^{-\Im m \chi(b,s)}\cos{\Re e\chi}]\\
\sigma_{el}=\int d^2{\vec b}|1-e^{i \chi(b,s)}|^2\\
\sigma_{inel}=\int d^2{\vec b}[1-e^{-2 \Im m {\chi}(b,s)}]
\label{stots}
\end{eqnarray}
The  introduction of  the jet cross-section as the term which drives the rise in
the eikonal function can be done unambiguously  through the inelastic cross-section, which only depends upon the imaginary part of the eikonal function. Notice that the expression for $\sigma_{inel}$
can also be obtained upon summing multiple
collisions which are Poisson distributed with an average number
$n(b,s)=2\ \Im m~{\chi}(b,s)$.
Using the experimental value of the $\rho$ parameter (the ratio of the real to the imaginary part of the forward elastic
amplitude) for $pp$ and $p\bar{p}$ processes, we can estimate only about $4\%$ correction from the real part. Hence,
we assume $\Re e \chi=0$, and  thus obtain a very simple approximate expression
\begin{equation}
\sigma_{tot}=2 \int d^2{\vec b}[1-e^{-n(b,s)/2}]
\end{equation}
which can be used to test our ideas about the underlying strong interaction dynamics.

\item the    impact parameter distribution, which contains all the b-dependence, represents the matter distribution in the colliding hadrons  and our model obtains it as the shadow of the path followed by the partons as they scatter through the hadronic matter. The scattering process defines the partonic densities in the transverse plane.
\item 
soft gluon emission
is a  QCD mechanism which introduces acollinearity and thus can reduce the LO parton-parton cross-section. 
 We use the resummation scheme obtained first in  QED by Yennie, Frautschi and Suura  \cite{YFS},
  extended to QCD transverse momentum ($k_t$) distributions   
  in refs. \cite{ddt,pp},   applied to Drell-Yan pair and W-production in \cite{halzenscott,altarelli}, as well as to jet production in \cite{CGS}. Earlier, we had discussed such $k_t$-resummation scheme for constant but large values of the coupling constant in \cite{ourkt}.
  As we discuss in Sect. 4, this resummation scheme
   includes
  an integration over the zero momentum modes.  Depending on the (yet unknown)  behaviour  of $\alpha_s$ in this region, the integral over the low energy part of  the single soft gluon distribution may become relevant.
  Similar integrals, involving $\alpha_s$ in the infrared region have been   discussed in \cite{dokshitzer} as  being related indeed to physical observables.
  For the transverse momentum case, this integral is usually absorbed in the intrinsic transverse momentum of the scattering hadrons and considered to be a constant. Our approach differs and has been discussed in \cite{lastPLB} and references therein. We shall return to this in Sect. \ref{bnsection}.
\end{enumerate}

The BN model was applied to proton-proton scattering, obtaining a total
cross-section at LHC $\sigma(\sqrt{s}=14\ TeV)= 100 \pm 12 \  mb$, where the error
reflects various uncertainties such as in the choice of  parton
densities  for the proton,
minimum parton $p_t $ cut-off, called $p_{tmin}$,  and the infrared behaviour of soft gluon coupling.
Thus, the model has a number of parameters, some of which have a physical
meaning associated with confinement.  As such, we do not know how and if to
change them as one goes from protons to photons. We shall try to vary them by
no more than (5-10)\% from their values for the proton. Whenever a stronger
variation is required, we shall comment upon it. The model predictions are
obviously dependent on the parton densities in the photon:  as in the case of
the proton, we have employed different  available sets,   obtained by
fits to the data on the photon structure function $F_2^{\gamma}$,
and seen how best to
  describe the available data   without appreciably changing
the parameters. Application  to photons however requires an
additional insight. In the eikonal representation we need to {\it adapt}
the hadronic language to that for the photon. One first needs the probability,  $P_{had}$, that
a photon behaves like a hadron and subsequently  one may then use the eikonal representation, as  in Refs.
\cite{aspen,sarcevic,halzen}:
\begin{equation}
\sigma^{\gamma p}_{tot} =
2 P_{had}\int d^2 {\vec b}[1-e^{-n^{\gamma p}(b,s)/2}].
\label{sigtot}
\end{equation}
Eq.~\ref{sigtot} has no 
{\it 
ab initio
} 
derivation. It  is  an   approximation, used in minijet inspired descriptions of  photon cross-sections,   useful  to model the low energy term and provide a normalization for the cross-section.
One could  expect  $P_{had}$ to have an energy dependence or b-dependence, or both. However, to  compare our model with other mini-jet models in the literature, and to make the simplest possible extension of our model from protons to photons, we consider it to be a constant as far as energy is concerned. As for the impact parameter space  and related energy dependence, these effects
are obtained in our model   through    QCD soft gluon resummation, just as in the case of protons.

In Eq.~\ref{sigtot}  the real part of the eikonal has been approximated to zero,  
again following the proton model, while
the
imaginary part is obtained from   the average number of inelastic collisions
for a given impact parameter $b$, $n^{\gamma p}(b,s)$, at a given c.m. energy $\sqrt{s}$.  Following our BN model for protons, we distinguish   between collisions calculable as QCD mini-jets, and  everything  else,  writing  the average number of collisions  as
\begin{eqnarray}
\label{nbs}
n^{\gamma p}(b,s)=n_{soft}^{\gamma p}(b,s) +n^{\gamma p}_{hard}(b,s)\nonumber \\ = n_{soft}^{\gamma p}(b,s) + A^{\gamma p}(b,s) \sigma_{jet}^{\gamma p}(s)/P_{had}
\end{eqnarray}
with  $n_{hard}$ including all outgoing parton processes with $p_t>p_{tmin}$.
 In Eq.~\ref{nbs}  the impact parameter dependence has been factored out,  averaging over densities in a manner similar to what was done for the case of the proton  in \cite{corsetti}. Because the jet cross-sections are calculated using actual  photon densities,
which themselves give the probability of finding a given quark or gluon in a
photon, $P_{had}$ needs to be  canceled out in $n_{hard}$.
 As for its value,    $P_{had} \approx P_{VMD}$.  $P_{had}$ is {\it not} the same numerical factor $R_{\gamma}$ used in Fig.~\ref{Fig:TX} to normalize all  the cross-sections at low energy, but  it can be connected to it  by making an expansion of the eikonal in the  low energy region, where $\sigma_{jet}\approx 0$, as shown at the end of this section.
 Also,  while $P_{had}$  can be factored out in some models, as we shall see later, this does not happen in the BN model.

The mini-jet cross-section is obtained by integrating the standard QCD
inclusive jet cross-section, using  a lower cutoff $p_{tmin}$ as described in
Appendix A.
The mini-jet cross-sections are  to be calculated using
parton densities (PDFs)  for the proton and photon determined from
perturbative QCD analysis of the data on $F_2^{p}, F_2^{\gamma}$ as
well as a variety of other data on hard processes for the proton.
Common ones for the proton are
GRV \cite{GRV},   MRST \cite{MRST}, CTEQ \cite{CTEQ}, whereas
those for the photon are GRV\cite{GRVPHO}, GRS \cite{GRS}, CJKL \cite{CJKL}.
These densities are available both at leading order (LO) or
higher, but in our model we use only the LO ones, as
part of the NLO effects are described by soft gluon
resummation and the use of NLO would result in some double counting.
Of course, in using  densities and parton-parton cross-sections only at LO but with   resummation of soft gluons, our model lacks  the non-infrared part of the  NLO corrections.
Since we consider the resummation effects in the infrared region  to be the most important for
saturation and these are easily incorporated in our model, we have opted for LO  densities, and thus also tree level parton-parton cross-sections and one loop $\alpha_s$.
We show in
Figure \ref{Fig:miniphoton_qmax94}
the energy
dependence of the mini-jet cross-sections for $\gamma p$ collisions,
for two different  sets of parton densities for the photon, GRS and CJKL.
We have used different values of the cut-off, namely
$p_{tmin}=1.2,1.3, 1.4\  GeV$ for GRS densities, higher values for the case of
CJKL densities, which give jet cross-sections rising faster with energy
than those calculated using GRS \cite{photon07}.  As for the proton densities, we have done all the
model calculations using  GRV94.
\begin{figure*}[htbp]
\centerline{
\hspace{-2cm}\includegraphics[scale=0.40]{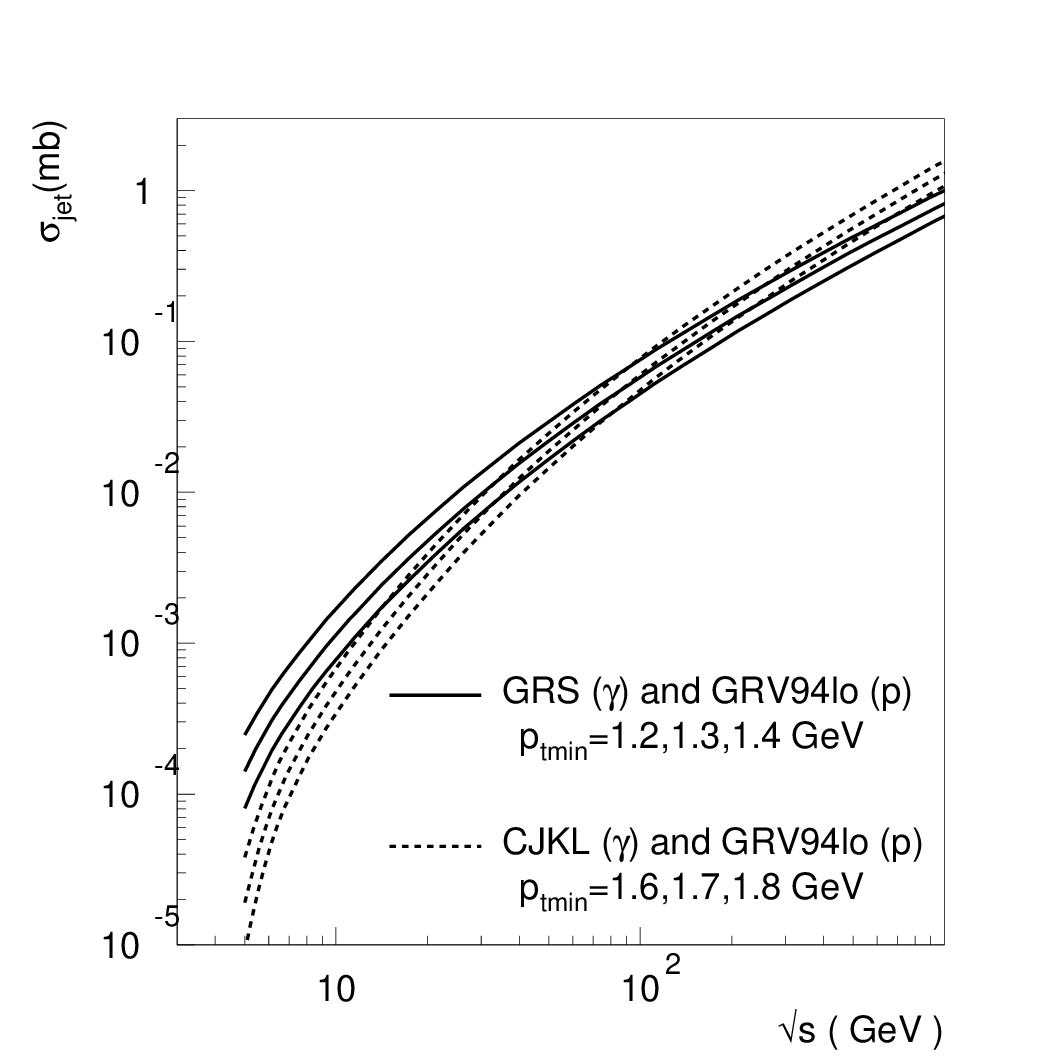}\hspace{.5cm}
\includegraphics[scale=0.40]{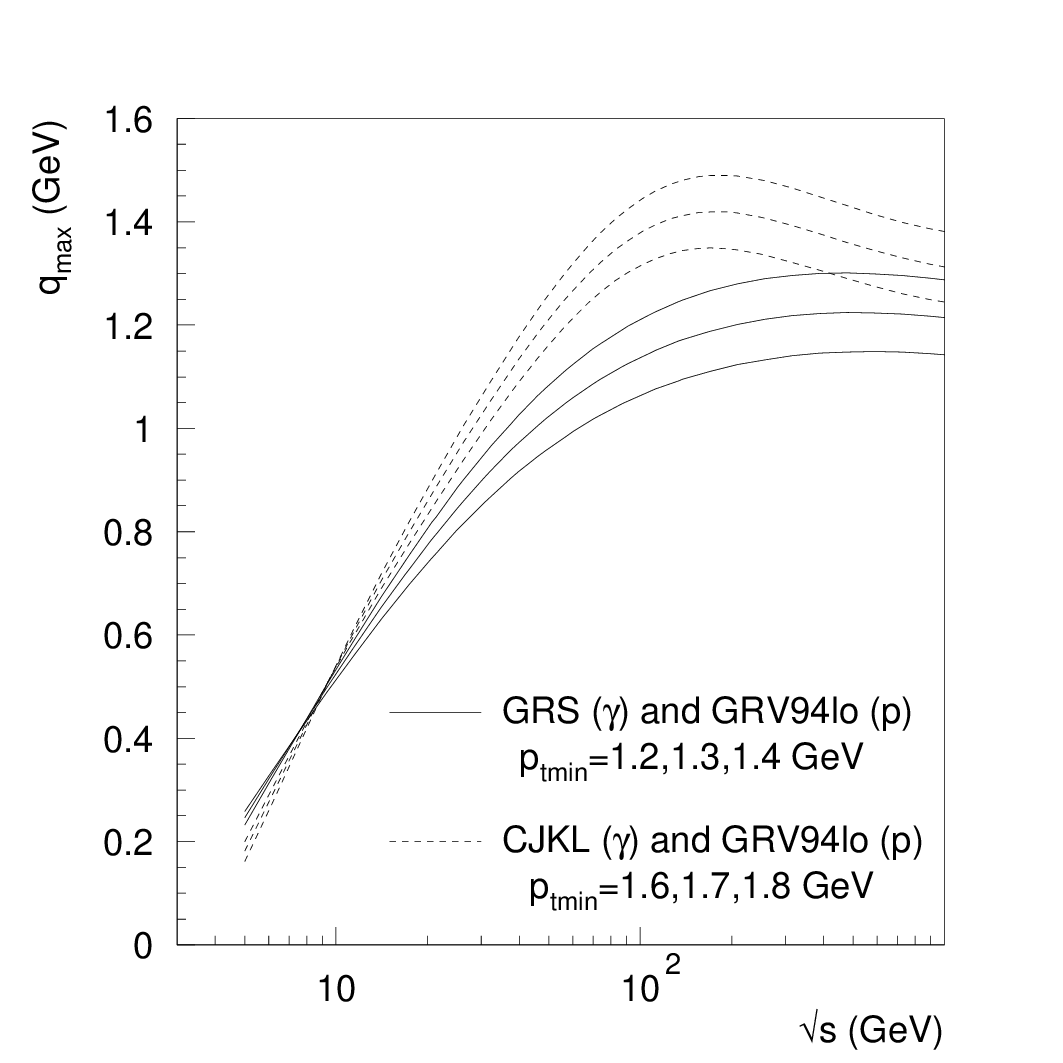} \hspace{0.1cm}}
\caption{Left panel: Photon-proton jet cross-sections for different densities
and  a range of  $p_{tmin}$ values. Right panel:  average value of the maximum  transverse momentum allowed for single initial state soft gluon emission, in $\gamma p$ scattering.}
\label{Fig:miniphoton_qmax94}
\end{figure*}

These cross-sections grow very rapidly with energy, reflecting
the infinite range of QCD. Since the finiteness of strong interactions
is reflected by the finite spatial extension of hadrons, one expects that
the eikonal representation would check such growth through the impact
parameter distribution which appears in Eq. \ref{sigtot}. A frequently used
distribution is obtained as a convolution of the form factors of the
colliding hadrons \cite{durand}, namely
\begin{equation}
A^{AB}_{FF}(b)=\int {{d^2 {\vec q} }\over{(2\pi)^2}}{\cal F}^A(q){\cal F}^B(q) e^{i {\vec q}\cdot {\vec b}}.
\end{equation}
However, it was noted already  in case of proton cross-section \cite{our99} ,  that,  without the inclusion of additional parameters,  this choice is unable
 to
reproduce both the early rise and the expected, Froissart -like,  subsequent
leveling off at high energies.
  Apart from this purely phenomenological consideration, the form factor description becomes undefined when dealing with photons. For photons, such models,
labeled as  Form Factor (FF) models, depend on  how one defines  the photon
form factor. In the literature, early  attempts to apply the mini-jet eikonalized expression to  the photon cross-sections  \cite{sarcevic} used  a monopole expression for  the photon (as  in  the pion case)  and the usual
 dipole expression for the proton form factor
with $\nu^2= 0.71\ GeV^2$, obtaining
\begin{eqnarray}
A^{\gamma p}_{FF}
&=&{{1}\over{4\pi}}
 {{
 \nu^2 k_0^2}\over{k_0^2-\nu^2
 }}
[
\nu b{\cal K}_1(\nu b) \nonumber \\
&-&
{{2\nu^2}\over{k_0^2-\nu^2}}
 (
 {\cal K}_0(\nu b)-{\cal K}_0(k_0b)
 )
]
\end{eqnarray}
with $k_0^2= 0.44\ GeV^2$ .  The above expression can be adapted to photon data  by varying the parameter $k_0$, and then the pion form factor expression for the photon  can be understood to represent an intrinsic transverse momentum  \cite{ourepjc,photon03}. In the Aspen
model \cite{aspen} another possibility has been explored. Namely, the overlap function is pa\-ra\-me\-tri\-zed as the Fourier transform of a dipole form factor
\begin{equation}
W(b,\mu) = {{\mu^2}\over{96\pi}}(\mu b)^3 K_3(\mu b)
\end{equation}
with three different scaling parameters for the three terms in which the eikonal is split, quark-quark, quark-gluon or gluon-gluon scattering. In this model, one then  uses
a single  functional expression for  the  b-distributions in hadron-hadron,  hadron-photon or photon-photon scattering, but  the difference  between these different processes is provided through the    $\mu $ parameters   which scale among the various processes according to the additive quark model. In another QCD inspired model\cite{luna}, a similar modelling has been made.

 More fundamental   attempts   to obtain the photon  impact factor in the context of perturbative QCD can be found in \cite{bartels} and references therein. 
 However, since these models are derived for virtual photon scattering, they cannot be used in the context of our BN mini-jet model, which is applied to real photon processes, with real photon parton densities. In    this paper we follow the same  strategy which we  used in case of the proton cross-sections and, 
for hard collisions,   use mini-jets and soft gluon resummation, with   $n_{hard}$ given by:
\begin{equation}
n_{hard}^{\gamma p}(b,s)={{
A_{BN}^{\gamma p}(b,s) \sigma_{jet}^{\gamma p}
 } \over{P_{had} }}
\end{equation}
with
\begin{eqnarray}
 A_{BN}^{\gamma p}(b,s) =
{\cal N} \int d^2 {\bf K}_{\perp} {{d^2P({\bf K}_\perp)}\over{d^2 {\bf K}_\perp}}
 e^{-i{\bf K}_\perp\cdot {\bf b}} \nonumber \\
 = {{e^{-h( b,q_{max})}}\over
 {\int d^2{\bf b} e^{-h(b,q_{max})} }}\equiv A_{BN}(b,q_{max}(s)).\ \ \
 \label{Eq:abn}
 \end{eqnarray}
 The function $A^{\gamma p}_{BN}$ is normalized to 1 and is obtained from
the Fourier transform  of the soft gluon resummed transverse
momentum distribution, whose structure we discuss in the next section.

 To complete the calculation of  $n_{hard}^{\gamma p}$, one has   to specify the value of  $P_{had}$, which in eikonal models \cite{aspen,halzen} indicates the probability that a photon behaves like a hadron  and is defined by the low energy part of the cross-section. At low energy, namely for $\sqrt{s}\approx 5\div 10 \ GeV$,  the mini-jet cross-section is 
 very small and $n(b,s)^{\gamma p} \approx n_{soft}^{\gamma p}(b,s)$. This part of the cross-section is outside the range of perturbative QCD we have described so far. Using Eq. \ref{sigtot}, we find that we can get a good description of the low energy $\gamma p$ data for the total cross-section with
\begin{equation}
n_{soft}^{\gamma p}(b,s)= {{2}\over{3}}  n_{soft}^{p p}(b,s)
\end{equation}
where $n_{soft}^{p p}(b,s)$ is the same function we have used for our
description of proton-proton collision in  ref. \cite{lastPLB} and $P_{had}=1/240 $,  a result consistent with Eq. ~\ref{rgamma}.

\section{\label{bnsection}The impact parameter distribution and the saturation parameters}
The distribution $A_{BN}$ is
energy dependent through the quantity $q_{max}(s)$, which represents the
average maximum transverse momentum allowed to a single soft gluon emitted  in the initial state  in a
given hadronic collision. This quantity is the input to the kernel
$h( b,q_{max})$,  which
describes the exponentiated, infrared safe, number of single soft gluons of
all allowed momenta  and is given by,
\begin{eqnarray}
h( b,q_{max}(s))  =
\frac{16}{3}\int_0^{q_{max}(s) }
{{dk_t}\over{k_t}}
 {{ \alpha_s(k_t^2) }\over{\pi}} \nonumber \\
\times \left(\log{{2q_{\max}(s)}\over{k_t}}\right)\left[1-J_0(k_tb)\right]
\label{hdb}
\end{eqnarray}
In the BN model, this function provides the cut-off in impact parameter space which softens the rapid rise of the mini-jet cross-section. To render this point clear, we shall summarize and outline our argument in what follows.

 The  soft gluon resummation formula in the
transverse momentum variable  has been known for a long time and reads \cite{ddt,pp,ourkt}:
\begin{equation}
d^2P({\bf K}_\perp)=d^2 {\bf K}_\perp
\int {{d^2 {\bf b}} \over {(2 \pi)^2}}
e^{-i{\bf K_\perp\cdot b} -h( b,q_{max}(s))}
\label{d2pk}
\end{equation}
with
\begin{equation}
h( b,q_{max}(s)) =\int_0^{q_{max}(s)}
  d^3{\bar n}(k) [1-e^{i{\bf k_t\cdot b}}]
\label{hdb1}
\end{equation}
where $q_{max}(s)$ is the maximum transverse momentum kinematically allowed for single emission.
The expression in Eq.~\ref{d2pk} can be obtained from the more general expression for soft resummation
\begin{equation}
d^4P( K)=d^4  K
\int {{d^4x} \over {(2 \pi)^4}}
e^{i{ K \cdot x} -h( x,E)}
\label{d4pk}
\end{equation}
with $h(x,E)$ similarly defined as in Eq.~\ref{hdb1}. This expression was obtained long time ago in QED \cite{YFS}, with order by order cancellation of the infrared divergence in perturbation theory. Notice that the same expression is obtained very simply in a semi-classical way using the methods of statistical mechanics and imposing energy momentum conservation \cite{ETP}. In such derivation, it is  energy-momentum conservation that brings in the cancellation of the infrared divergence between soft and real quanta emission.

In QED,  $d^3{\bar n}(k)\propto \alpha \log ( {{ 2q_{max} } \over{
m_{electron} }})$ and resummation in the transverse momentum variable
is well approximated by a  first order expansion in $\alpha$. For large values of the coupling constant however,  this approximation would be inadequate and we  noticed long time ago \cite{ourkt} that, for such cases,   resummation of  soft quanta emission can provide a transverse momentum cutoff, based on the expansion of the Bessel function $J_0(k_tb)$ around $k_tb\approx 0$. 

 In  QCD
\cite{ddt,pp},  Eq.~\ref{hdb1} 
shows the presence of  a non-trivial complication, namely the impossibility to use the asymptotic freedom expression for $\alpha_s$ down to $k_t=0$.
 It is this non-trivial complication  which we exploit to study scattering in the very large impact parameter region.

 To overcome the difficulty arising from the infrared region, the function $h(b,E)$, which  describes the relative  transverse momentum distribution induced by soft gluon emission from  a pair of,
initially collinear, colliding partons  at LO, is split into
\begin{equation}
\label{h1}
h(b,E) = c_0(\mu,b,E)+ \Delta h(b,E),
\end{equation}
where
\begin{equation}
\label{h2}
\Delta h(b,E) =
 \frac{16}{3} \int_\mu^E {\alpha_s(k_t)\over{\pi}}[1- J_o(bk_t)]
 {{dk_t}\over{k_t}}
  \ln {
  {{2E}\over{k_t}}}.
\end{equation}
Since, in $\Delta h(b,E)$, the integration only extends down to the  scale $\mu$ (not zero),
the $J_o(bk_t)$ is assumed to oscillate to zero and hence is dropped.
The last integral is now independent
of $b$ and  can be performed, giving
\begin{eqnarray}
\Delta h(b,E) =
\frac{32}{33-2N_f}\nonumber \\
\bigg\{ \ln ( {\frac{{2E}}
{\Lambda }} )\left[ {\ln ({\ln ( {\frac{E} {\Lambda }})}) -
 \ln
({\ln ( {\frac{\mu } {\Lambda }} )})} \right]
- \ln ( {\frac{E}{\mu }} ) \bigg\}. \ \ \ \ \
\end{eqnarray}
where $\Lambda$ being the scale in the asymptotic freedom expression for $\alpha_s$.
In the range $1/E < b < 1/\Lambda$
the effective $h_{eff}(b,E)$
is obtained by setting $\mu = 1/b$ 
\cite{pp}. This choice of the scale   introduces a cut-off in impact parameter space which is stronger than any power, since
the radiation function  is now
 \begin{equation}
e^{-h_{eff}(b,E)} =
 \big{[}
{{
ln(1/b^2\Lambda^2)
}\over{
ln(E^2/\Lambda^2)
}}
\big{]}^{(16/25)ln(E^2/\Lambda^2)}
\label{PP}
\end{equation}
which is Equation(3.6) of ref. \cite{pp}. The remaining  b-dependent terms in $h(b,E)$ are dropped, a reasonable approximation if one  assumes that there is no physical singularity in the range of integration
 $0\le k_t \le 1/b$.
 However, when the integration in impact parameter space extends to very large b-values, as is the case for the calculation of total cross-sections, this expression fails  to reproduce the entire range of the energy dependence of low energy
transverse momentum effects. To explore the very large b-region, we suggest to use Eq. (\ref{hdb}) with its full integration range, pro\-po\-sing a phenomenological approach to the zero-momentum soft gluons
\cite{ptintrinsic}. Our  approach uses a singular, but integrable expression for $\alpha_s$:  this allows us to extend the integral to the minimum allowed value zero  and to obtain a b-dependence which is stronger than the one of Eq.~\ref{PP}.

Thus the infrared   region can provide an impact parameter  cutoff at large b-values, provided the integral is finite. This requires any proposed expression for $\alpha_s$ in the infrared region  to  be integrable. As discussed  in detail in ref. \cite{our99}, the actual functional dependence of the cut-off depends on the model for $\alpha_s$ in the infrared. Here we  mention that in the model we propose, the  resulting  cut off in b-space is at least an exponential function.

In the next
subsections, we shall
discuss various  proposals to model the infrared region and   how the full  integral of Eq. (\ref{hdb})  controls the
saturation of the cross-section through its limits of integration. We shall see that  
  our model for  soft gluon emission   is regulated by a constant
infrared parameter $p$ and the energy dependent momentum function $\qmax $
as follows:
\begin{enumerate}
\item the energy dependent momentum saturation parameter $\qmax(s)$
depends on the energy behaviour of the  density functions of colliding partons  and on $p_{tmin}$, the mini-jet cut-off,
\item the infrared parameter \p , to be specified short\-ly,  defines the infrared behaviour of $\alpha_s(k_t^2)$.
The closer its value  is to 1,  the more the mini-jet
cross-sections will be quenched at any given energy.
\end{enumerate}

\subsection{The  momentum saturation parameter $q_{max}(s)$}
For any given parton parton collision, $q_{max}(s) $ can be defined by
kinematics. We introduced  this quantity for the first time in
\cite{corsetti} to represent the maximum transverse momentum carried  by
a single gluon,  averaged over the basic scattering cross-section with a
procedure  described in Appendix B  for the convenience of the reader.
To highlight the physical meaning of $\qmax(s)$,
let us  define  the saturation  parameter
\begin{equation}
{\hat \kappa}= {{\sqrt{\hat s}-\sqrt{{\hat s}_{jets}}
} \over { {\sqrt{\hat s} }/2 }}
\end{equation}
for each parton pair of c.m.  sub energy \shat \  which scatters into a
final parton pair of c.m. energy
$\sqrt{{\hat s_{jets}}}$.  Let us now use the kinematics of the process
\begin{equation}
parton(x_1)+parton(x_2)\rightarrow gluon(k_t) + jet_1+jet_2
\end{equation}
to write  the maximum transverse momentum of the emitted gluon, in the case
of limited energy loss as
\cite{mario}
\begin{equation}
k_{tmax}={{\sqrt{\hat s}}
\over{2}}
(1-
{{
{\hat s}_{jets}
}
\over
{
{\hat s}
}}
 )
 \approx
{{
\sqrt{\hat s}
}\over{2 }}
 {\hat \kappa}
\label{Eq:ktmax}
\end{equation}

This quantity plays an important role in our model.  As the available
c.m. energy increases, it increases, depending upon the
pro\-ba\-bility of producing a  parton pair  scattering into a
given final state. It thus depends upon the densities and the parton-parton
cross-section. As it increases, more and more a\-col\-linearity is introduced
in the scattering and the stronger is then the reduction in the growth of the
mini-jet cross-section.

Notice that now there appear  two  different scales and both low-x perturbative  gluons
as well as soft gluons. We stress the distinction between them: low-x gluons
participate in the hard parton-parton scattering described by the mini-jet
cross-section discussed in the previous section, for which
\begin{equation}
p_{tout}\equiv p_t^{jet}\ge \ptmin\approx 1\div 2 \ GeV
\end{equation}
 These low-x perturbative gluons   interact  with a strength proportional to
$\alpha_s(p^2_{tout})$, while soft gluons are those
emitted, from the initial state, in any given parton-parton process with  transverse momentum
\begin{equation}
k_t\le \ktmax  \approx 10  \div 20\% \  p_{tout}
\end{equation}
 This scale, $k_{tmax}$ defines the single soft gluons, whose number  can be indefinite.
These soft gluons need to be re-summed through the procedure which results in the exponentiated factor of Eq.~\ref{Eq:abn}.

In a model such as ours,
 which is not a Monte Carlo simulation of the processes involved,
we have opted for
 averaging  these effects,   embodying them in a
factorized expression such as that given by Eq. \ref{Eq:abn},  with $\ktmax$
averaged out to obtain $\qmax$, as shown in Appendix B.  The expression for
$q_{max}(s)$ depends both on the parton densities and the value of $p_{tmin}$.
The resulting quantity is energy dependent since the densities are energy
dependent through  the applied DGLAP evolution.
The averaging process done in this model  includes only quark densities as
 the source of the leading  acollinearity effect.
We  consider  the leading effect to arise because of soft gluon emission from
the external legs  of the scattering process, valence quarks for the proton beam and all flavours of quarks for the photons.  An improvement of the model could include soft gluon emission also from the low-x perturbative gluons, as we shall discuss in a forthcoming paper.   In the right-hand panel of Fig. \ref{Fig:miniphoton_qmax94} we  show the  dependence of $q_{max}(s)$ upon the c.m. energy of the colliding
particles, for the same densities and $p_{tmin}$ values used in the mini-jet
cross-sections shown  in the left panel.

As $\qmax$ increases with energy,  the growth  of the total cross-section due to  mini jets  is tempered by
soft gluon emission,  through the exponential damping factor $e^{-h(b,q_{max})}$.  However, there is an equilibrium between the increase of
 $\qmax$ and the rate of increase of the mini-jet cross-section since one
reflects the quarks  and the other the gluon densities. The distribution
of these partons at high energy follows the parton sum rules and one is not
independent of the other.  From the right hand panel of
Fig. \ref{Fig:miniphoton_qmax94} we see that $\qmax$,  for both GRS and CJKL
densities, will reach some sort of saturation at high energies, which reflects
in the total cross-sections reaching a stable slope.

The momentum saturation parameter   $\qmax$  is not the only quantity
which gives rise to saturation, the infrared limit of $\alpha_s$ also
plays a major role. We shall discuss this in the next subsection.

\subsection{A phenomenological approach to the infrared limit of $\alpha_s$}
To complete the calculation of the impact parameter distribution for hard
processes in  $\gamma p$ collisions, we need to
discuss the lower limit of integration in Eq.\ref{hdb}. Usually, the soft
gluon resummation formula extends the soft gluon momenta to an   infrared cut-off
taken to
correspond to the intrinsic transverse momentum scale of the scattering
hadrons \cite{ddt,pp}. Instead,  in our model, we extend the integration down to the zero momentum modes.
 
The reason to do so lies in the nature of the cancellation of the infrared divergence. To obtain this cancellation, it is mandatory that virtual and real soft gluon emission join in the zero momentum limit. Of course we do not know how to deal with $\alpha_s$ in this limit and for this  reason  the integral over this region  is usually left out and substituted with a constant intrinsic transverse momentum. However, this part of the  integral is  relevant for many   minimum bias processes, as pointed out in ref. \cite{dokshitzer}. With our model, we aim to relate   the  behaviour of  the impact factor at very large b-values with the infrared region of  soft gluon emission. 

To do so, we need therefore to make an ans\"atz as to  the behaviour of the
strong coupling constant in the infrared region,
where the
usual asymptotic freedom expression for $\alpha_s(Q^2)$ cannot be used.
One possibility is to use an expression which would go to a constant
as  $Q^2 \rightarrow 0$ as in
\begin{equation}
\alpha_s(Q^2)={{12 \pi}\over{33-2N_f}} {{1}\over{\log[a+{{ Q^2}\over{\Lambda^2 }}]
 }}
\end{equation}
with $a\approx 2$ \cite{yndurain,halzenscott}
 and $\Lambda=\Lambda_{QCD}$. This expression is often referred to as the {\it frozen $\alpha_s$ } case. Another possibility is to employ the Richardson potential for quarkonium, which uses  a singular $\alpha_s$, namely   $a=1$, so that
 \begin{equation}
\alpha_s^R(Q^2)\approx {{1}\over{Q^2}} \ \ \ \ \ \ Q^2\rightarrow 0
\end{equation}
 The Richardson potential has been shown to give good results to describe charmonium states \cite{richardson}, but it cannot be used here because the integral over the soft gluon modes would diverge. The reason it works in quarkonium applications is that in that case  one never actually reaches values corresponding to $Q^2=0$, since the potential binds the two quarks in a
  region of space at fixed finite distance of ${\cal O} (r_{Bohr})$.

 In order to be able to use  the Richardson-like $\alpha_s^R$, we soften the singularity with the  proposal  that in the infrared limit, one can phenomenologically
use the expression
\begin{equation}
\alpha_s(k_t)= constant \times \left({{\Lambda}\over{k_t}}\right)^{2p}\ \ \ \ k_t\to 0
\label{phenoas}
\end{equation}
where $\Lambda$ is a cut-off of order $\Lambda_{QCD}$,  and $p$ is a parameter which  embodies
the infrared behavior, with  $p<1$ so that the soft gluon integrals
converge.  For the time being, we consider  the above expression
as a phenomenological ans\"atz .  The constant in front of Eq.~\ref{phenoas} should be chosen to
provide a smooth extrapolation to the perturbative expression for
$\alpha_s$. Our choice for the interpolating function is
\begin{equation}
\alpha_s={{12 \pi}\over{33-2N_f}} {{p}
\over{\ln[1+p({{k_t}\over{\Lambda}})^{2p}]}}
\label{alphas}
\end{equation}
This expression was also introduced to describe the intrinsic transverse momentum of Drell-Yan pairs, with the choice
$\Lambda =100\  MeV$ \cite{ptintrinsic}  and $p=5/6$.
This choice for the infrared behaviour (zero momentum gluons) was motivated \cite{ourlast} by
 an argument due to Polyakov \cite {polyakov}.
It is clear that the closer $p$ is to 1, the bigger the soft gluon integral $h(b,q_{max}(s))$ is and the stronger the saturation effects will be.

We shall show the results for the total $\gamma p$ cross section for this and  other models in the next section.

\section{Total $\gamma p$ cross-section at accelerator energies}

In this section we examine the data on $\sigma_{\gamma p}^{\rm tot}$,
starting from the low energy photoproduction
data \cite{PDG} up to the high energy
HERA data and compare them with model predictions.
In fact, HERA studied $ep$ scattering and not only yielded
information on energy dependence of the photoproduction
cross-sections but also on $\gamma^* p$ cross-sections for different
values of virtuality (invariant mass $-Q^2$) of the photon. For small
values of $x_\gamma$ and $Q^2$,  the measurement of
$F_2(W^2,Q^2)/Q^2$, where $W$ is the invariant mass of the $\gamma^*
p$ system,  gives a measurement of $\sigma^{\gamma^* p}$
through the relation,
$$
\sigma^{\gamma^* p} \simeq {4 \pi^2 \alpha \over Q^2} F_2(W^2, Q^2)
$$
The extrapolation of the measured quantity on the right hand side can then  give  the photoproduction cross-section
in the limit $Q^2_\gamma = 0$  \cite{dieter}.
  The extrapolation of data collected with the ZEUS Beam Pipe Calorimeter (BPC) \cite{bpc}, based on a Generalized Vector Meson Dominance model,  produces \cite{bernd,bpcdata} a set of measurements  in a continuous energy range $W_{\gamma p}= 104 \div 251\ GeV$  consistent within errors with the  photoproduction data. The systematic errors  are due to the GVMD extrapolation, with the longitudinal cross-section $\sigma_L=0$, of the ZEUS BPC95 data.
 In the comparison, we have  also included some cosmic ray data \cite{vereshkov} which are in an  energy range lower than HERA data, but higher than  the older photoproduction data.

 Let us start with the BN model for photons as described in the previous section.
We have used GRV densities for the protons \cite{GRV}  and have varied the photon densities, using
both GRS and CJKL. We show the result of the model and the dependence upon the model parameters in
Figs. \ref{Fig:grs_cjkl},\ref{Fig:cjklp}. In Fig.\ref{Fig:grs_cjkl} we have  varied $p_{tmin}$ and the
densities to describe the  high energy data from HERA in addition to the most acceptable description of
the beginning of the rise,  while keeping the parameter $p$ in a range close to the \pppbarp \ case.  In Fig. \ref{Fig:cjklp} we have allowed for a larger variation in  the value of the infrared parameter $p$, fixing the  PDF set and a range of  appropriate  values for $p_{tmin}$.

 In order to obtain a good model description,  we shall focus not  only on the HERA data, but also on  the beginning of the rise, as this signals the onset of  the contribution
of QCD processes and is strongly dependent upon $p_{tmin}$. We can see from  Fig. \ref{Fig:miniphoton_qmax94} that, for the range of $p_{tmin}$ values of interest, the  mini-jet  cross-sections calculated
with CJKL densities  rise faster   than those calculated with  GRS. It follows that, to describe the same HERA data, one will need to use different  values of $p_{tmin}$ depending upon the PDF set used.
Thus   CJKL densities  call for a
larger $p_{tmin}$ than  GRS densities.  In Fig. \ref{Fig:grs_cjkl}
 the infrared parameter $p$ has
been kept close to the value determined from the \pppbarp \ cross-section, namely $p\approx 0.7\div 0.8$.
  We see that  the range of acceptable $p_{tmin}$ values for GRS densities
 is not far from those used in the \pppbarp \ case, where $\ptmin \approx 1.1\div 1.25 \  GeV$,  but it is
higher for CJKL.

\begin{figure}[htbp] 
   \centering
   \includegraphics[scale=0.40]{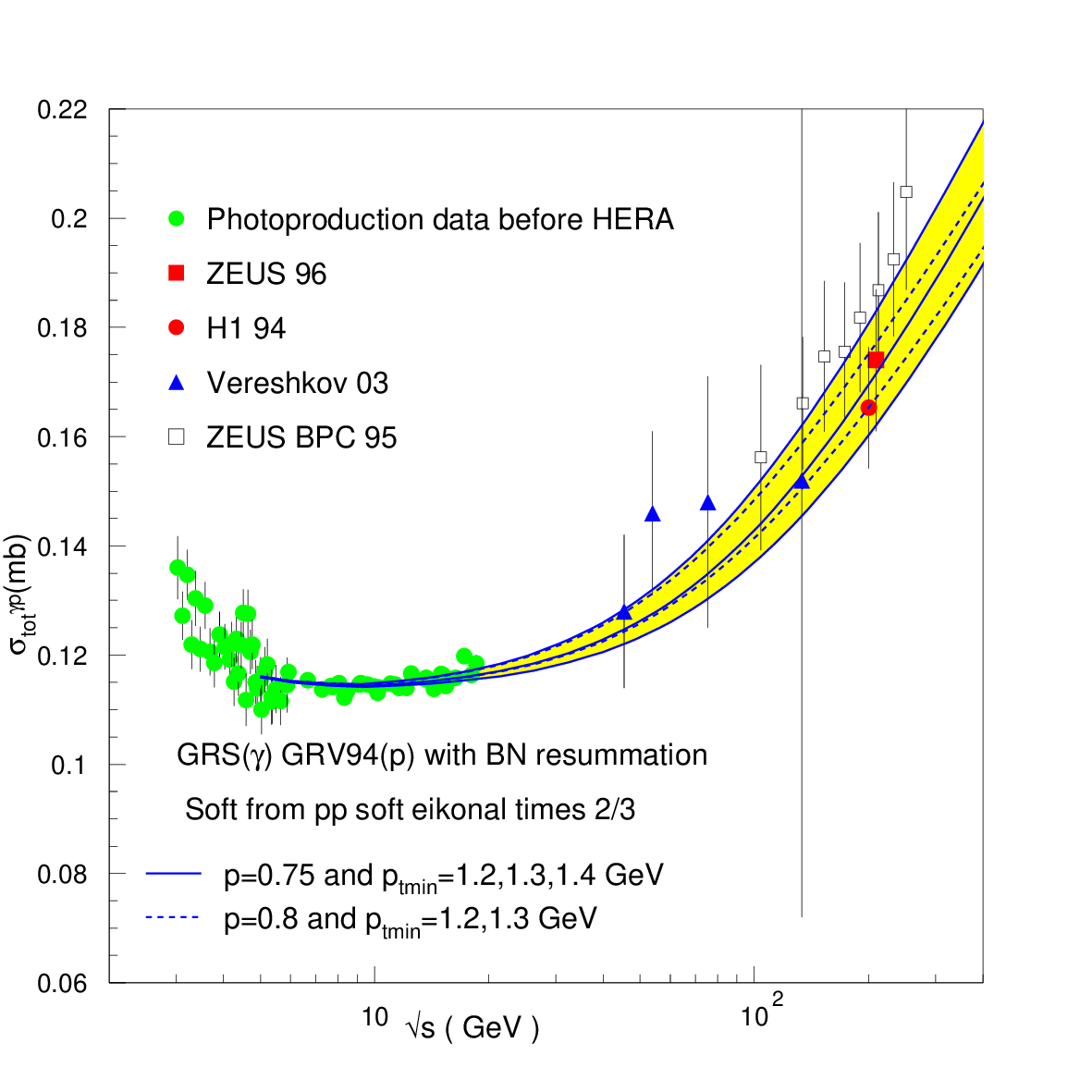}\hspace{.1cm}
\includegraphics[scale=0.40]{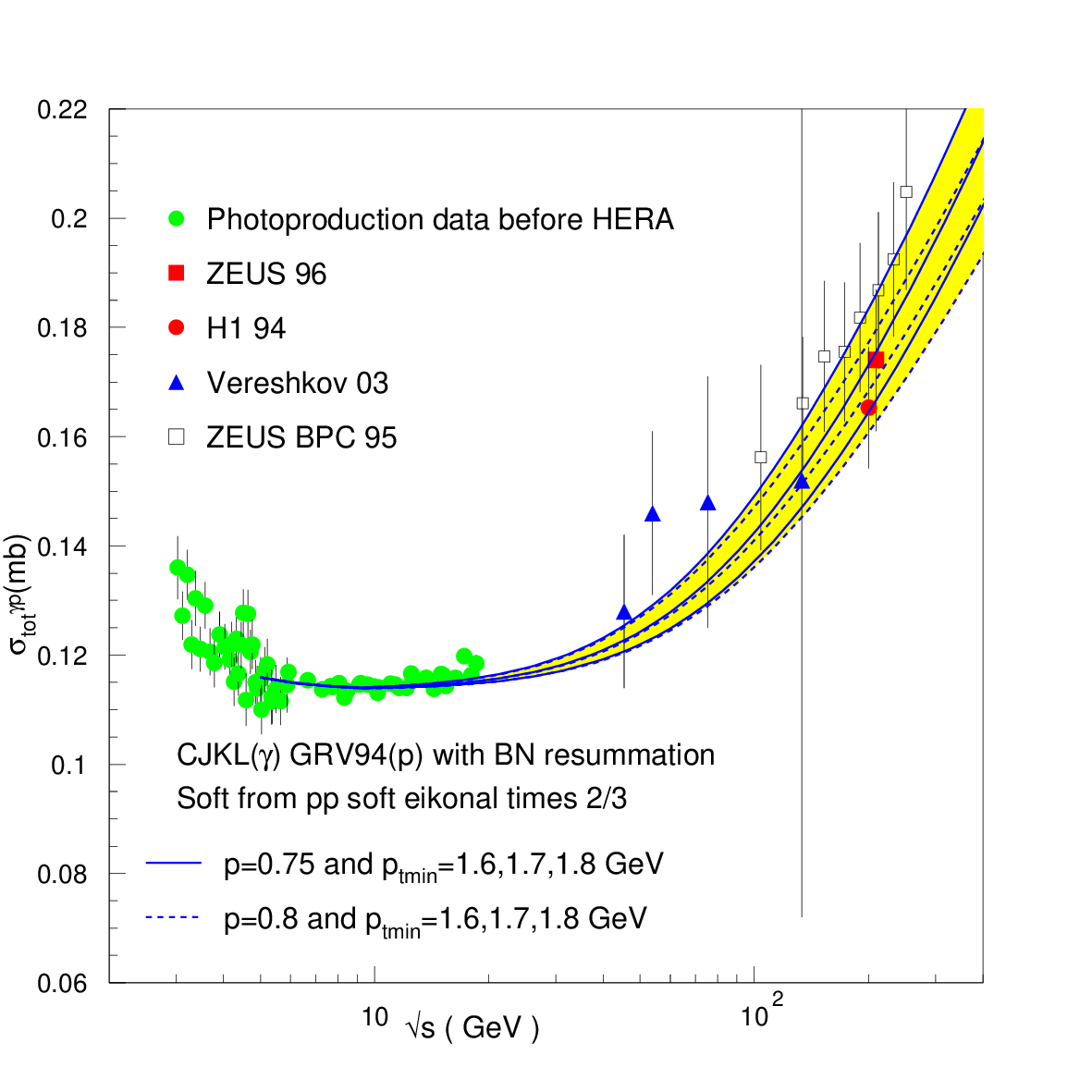} \hspace {.1cm}
   \vspace{0.2cm}
   \caption{Total $\gamma p$ cross-section with a range of parameter values close to the proton case,
 GRV densities for the proton and GRS or CJKL densities for the photon. Data from HERA
 are from Zeus \cite{zeus}, H1 \cite{h1} and a set of data from the ZEUS BPC extrapolated from $Q^2\ne 0$  \cite{dieter,bernd,bpcdata}.}
   \label{Fig:grs_cjkl}
\end{figure}
To summarize the results of these figures,  the latest HERA data are well described for a range of parameters
 $p=0.75\div 0.8$ and $p_{tmin}=1.2\div 1.3 \  GeV$, to be compared with the  $pp$ and
 ${\bar p} p$  case where the range was very similar, with our central value $p=0.75$ and
 $p_{tmin}=1.15\ GeV$ for GRV densities. A good description is also obtained with
CJKL densities, but then one needs a different \{$p,p_{tmin}$\} set, as one reads
from the second panel in Fig. \ref{Fig:grs_cjkl}.

\subsection{Parameter dependence}
In the previous figures, we have applied the BN model for protons to $\gamma p$ scattering. No attempt has been made to choose one particular curve through an evaluation of the $\chi^2$ of our curves with respect to the data. This is so because the aim of this paper is to extend the BN  model to photon processes with as little changes as possible from the proton case. Another reason is that   the data have rather large errors, and are extracted  from very different experimental situations, which range from cosmic rays to HERA data on photoproduction to extrapolations of data collected with the ZEUS BPC. We did not wish to select one particular data set and adopted the strategy to see whether  the range of parameter values used for  the proton results give curves consistent with the photo-production data.

The high energy behaviour of the model depends on the following entries:
\begin{enumerate}
\item PDF  set, 
for which
we have used GRV for the proton and  GRS or CJKL for the photon
\item minimum hard parton cutoff $p_{tmin}$
\item infrared parameter $p$
\end{enumerate}
with the choice of $p_{tmin}$ related to the chosen PDF set. Of the  two parameters, $p_{tmin}$ is basically fixed so as to reproduce the early rise, where soft gluon resummation is not yet important,   while the parameter $p$ controls the quenching of the rise at high energy and also the absolute value of  $n_{hard}(b,s)$. We also use  $p_{tmin}\approx 1\div 2\ GeV$ for the perturbative mini-jet calculation to make sense.  The dependence on densities and $p_{tmin}$ was shown in
Fig. \ref{Fig:grs_cjkl}, while the parameter $p$ was kept within the range of values used for the proton case. We see that the GRV/GRS density set is the one for which the extrapolation of the model from protons to photons works better. On the other hand, the model is also able to describe the data using CJKL densities, but in such case, the parameters are different from the proton scattering case.
In  Fig. \ref{Fig:cjklp}
we give examples of how it is possible to quench the   high energy rise from a given choice of PDFs (CJKL, in this example) through the parameter $p$.
\begin{figure}[htbp] 
   \centering
 \includegraphics[scale=0.4]{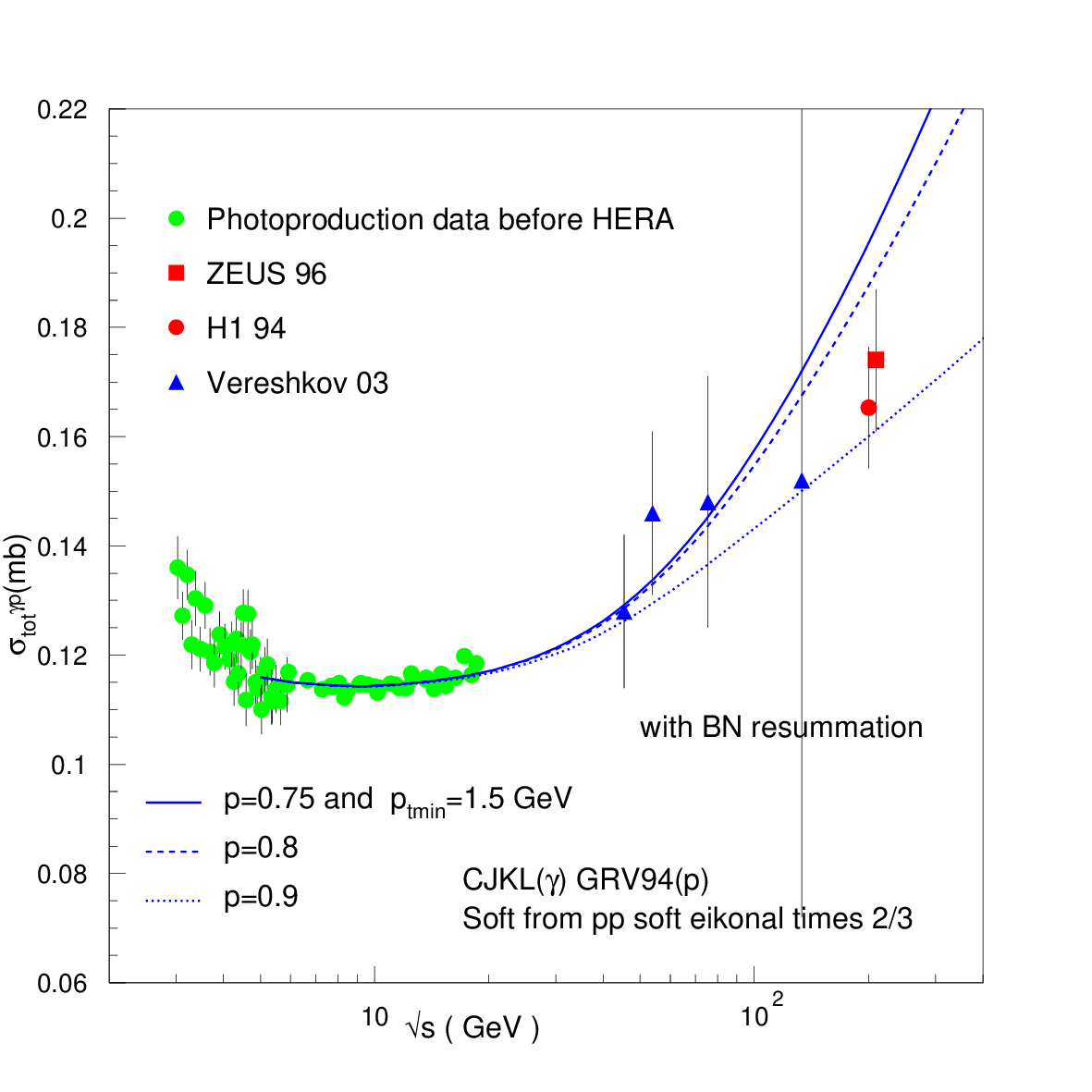}
   \caption{Total $\gamma p$ cross-section with  GRV for proton and CJKL densities for the photon, for
   a spread of $p$ values.}
   \label{Fig:cjklp}
\end{figure}
Notice that to catch the early rise, around $\sqrt{s}=20\ GeV$, for CJKL densities  one needs    $p_{tmin}\approx 1.5\ GeV$, but then this requires a larger $p$-value in order to quench the rise and not overshoot the HERA data points.

All in all, we can say that the model   adequately describes the photon-proton cross-section data
and  we can try to extend it to higher energies so as to make predictions for cosmic ray energies to be reached by the AUGER experiment \cite{zas,auger}. We turn to this problem in the next section.
 But before this, we address the question of factorization: is a photon like a proton just multiplied by a constant factor?
 From what we have seen so far, one could describe $\gamma p$ total cross-section up to HERA energies  either through a microscopic  model such as our BN model, with quarks and gluons,    or
 through other approaches based on various forms of factorization. In particular, the Aspen model also gives a good description as  do other approaches, based on multiplying the result of fitting \pppbarp \ data with a constant factor.   We shall discuss this point in the coming subsection.

\subsection{Factorization: a hadron-like  photon}
In the previous section, we have applied our model to the total
$\gamma p$ cross-section, using available photon densities, going through
the various steps defining our model, namely calculation of mini-jet
cross-sections, evaluation of the energy dependence saturation parameters,
determination of the energy dependent impact parameter function from soft
gluon resummation $A_{BN}(b,\qmax(s))$ and finally eikonalization.  In this
approach, at high energy, the photon is an independent entity from a hadron,
with the rising behaviour of the cross-section and the b-distribution of the
$\gamma p $ collision determined independently
from other $hadron-hadron$ collisions such as pp. This is different
from other models, for instance the Aspen model \cite{aspen}, where the
photon properties are obtained through scaling factors inspired by the
additive quark model. As a consequence, in the Aspen model for photons,
one can prove  a factorization property \cite{martinfacto} which would then
allow to extract the $\gamma \gamma $ cross-section simply as \cite{gribov}
\begin{equation}
\sigma_{tot}^{\gamma \gamma}
= {{
(\sigma_{tot}^{\gamma n})^2
}\over{ \sigma^{nn}_{tot} }}
\label{gribovfact}
\end{equation}
with $\sigma_n$ to indicate the nucleon cross-sections. We shall discuss
the $\gamma \gamma$ cross-sections within our BN model for photons in a
separate paper, however we notice that such factorization is not to be
expected in the model we present here.

Other types of factorization models are based on the Regge-Pomeron exchange,
keeping a constant universal behaviour of the rising part of the cross-section
with coefficients based on the factorization of the residues at the poles in
the elastic amplitude, so that
\begin{eqnarray}
\sigma_{tot}^{nn}=X_{nn}s^{-\eta}+Y_{nn}s^{\epsilon}\\
\sigma_{tot}^{\gamma n}=X_{\gamma n}s^{-\eta}+Y_{\gamma n}s^{\epsilon}\\
\sigma_{tot}^{\gamma \gamma}={{(X_{\gamma n})^2}\over{X_{nn}
}} s^{-\eta}+{{
(Y_{\gamma n})^2}
\over{Y_{nn}
}}s^{\epsilon}
\label{dlgamma}
\end{eqnarray}
with $\epsilon \approx 0.08\div 0.09$. This type of factorization is of
course different from the one in Eq.\ref{gribovfact}, but it still implies the
 idea that there is a universal behaviour of the energy dependence, not only
at low energy, where one can confidently assume that the hadronic interactions
of the photons are those of a vector meson, but also at  high energy.

Such a description of the photon, i.e, that the photon  is always hadron-like,
 could be reflected in our model by simply scaling the BN cross-section for
protons, as
 \begin{equation}
\sigma_{tot}^{\gamma p}=R_\gamma \sigma_{tot}^{pp}=R_\gamma \ 2 \int d^2 {\vec b}[1-e^{-n^{pp}(b,s)/2}]
\label{BNfacto}
\end{equation}
Present accelerator data for $\gamma p$ are consistent with factorization
models, including an application as given in Eq. \ref{BNfacto},  but as we
shall see in the next section, at higher energies, expectations will differ.

\section{Extrapolation to very high energies }
  In this section we extend our calculation beyond present accelerator
energies and compare  our predictions with other approaches.
We start  with the simplest factorization model of Eq. \ref{BNfacto} and  multiply the band of results obtained in ref. \cite{lastPLB} for proton-proton total cross-section  with a constant factor. This is similar to what we did in Fig. \ref{Fig:TX}, except that we use the full band from Fig. 2 of ref.\cite{lastPLB}. Let us indicate these predictions as  $BN_F=BN_{protons}/330$ (F for factorization).  We then compare this band with the results obtained using the BN model with photon densities,  GRS and CJKL, namely the curves shown in Fig. \ref{Fig:grs_cjkl}, extended to $\sqrt{s_{\gamma p}}=20\ TeV$.  This comparison is shown in
Fig. \ref{Fig:grs_cjkl20K}. We see that, at  energies  around and through  the TeV region,  the band  obtained from $\sigma_{tot}^{pp}$ falls short of what the BN model for photons ($BN_\gamma$) predicts. Other models, which enjoy factorization like the Aspen model, also remain lower than our curves.  While at moderate, HERA like energies, all the
three models, Aspen, $BN_\gamma$  or    $BN_F$
 give  acceptable fits to the data, there is a difference of almost   50\% among
their high energy extrapolations.  Thus, the first interesting conclusion from this
exploration of the very high energy region is that
there is a distinct difference between  predictions  from  our BN model and those  from the
QCD inspired model of Block et al. (Aspen) \cite{aspen}, as well as from a straightforward
multiplication of  our band of predictions for the proton times a
normalization factor.
\begin{figure}[htbp] 
   \centering
   \includegraphics[scale=0.4]{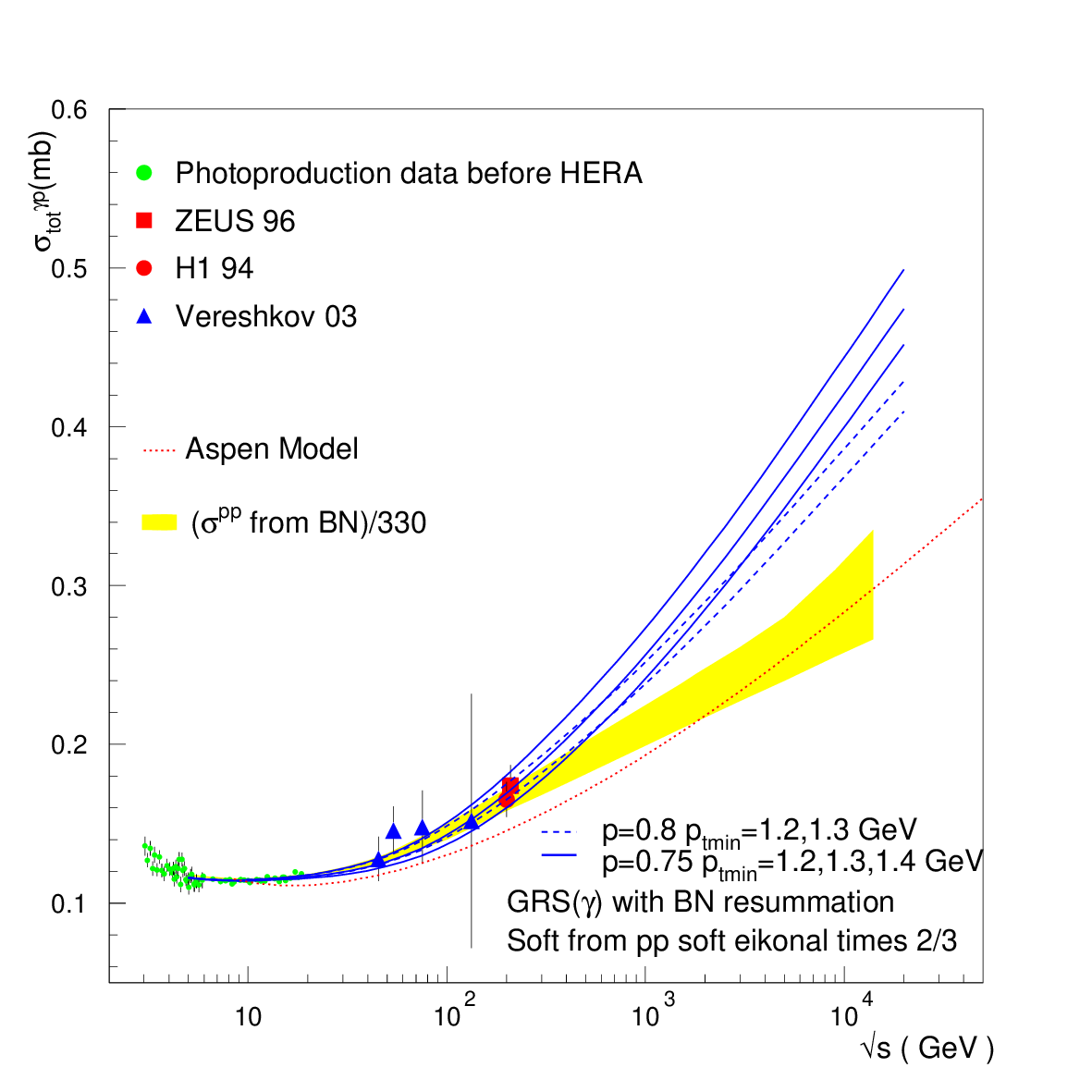}\hspace{0.1cm}
 \includegraphics[scale=0.4]{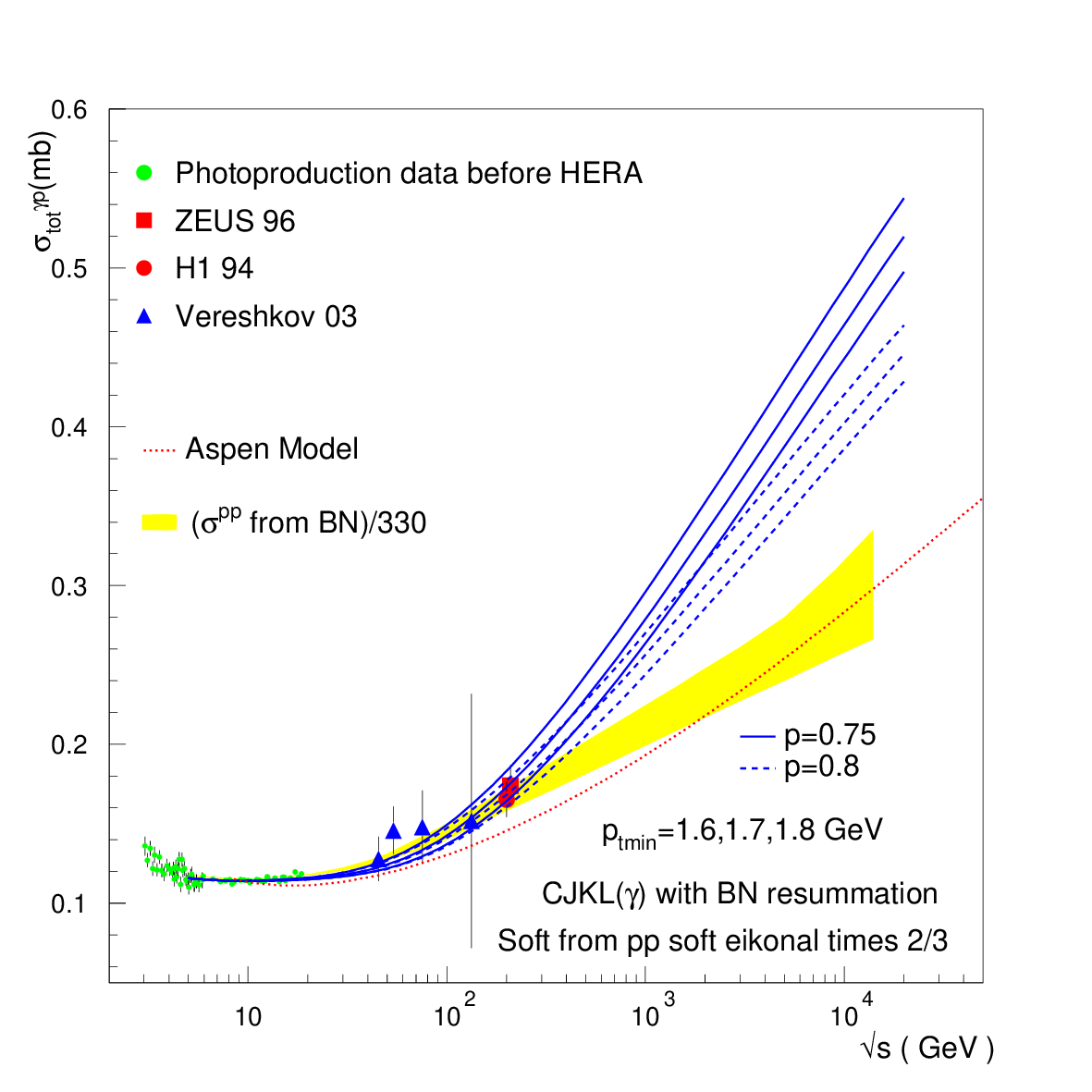} \hspace {.1cm}
\vspace{0.2cm}
   \caption{Total $\gamma p$ cross-section with GRS and CJKL densities, compared with
ref. \cite{aspen} predictions and with a brute force
factorization of our proton-proton results from \cite{lastPLB}. }
   \label{Fig:grs_cjkl20K}
\end{figure}

The next interesting result from this extrapolation appears when one compares our
 model predictions with the fit to HERA data  performed  by Block and Halzen  and  based on a low
energy parametrization
of $\gamma p$ resonances joined with Finite Energy Sum Rules (FESR) and asymptotic $\ln^2{s}$ behaviour
\cite{bhfit}.
 Fig. \ref{Fig:comparison}
shows
a band corresponding to the predictions of our model for photons (upper band)
compared to
$BN_F$ (lower band), the  Block and Halzen fit \cite{bhfit}, the Aspen model of \cite{aspen}, and  an
eikonal mini-jet curve
 which uses the proton and pion form factors for the
impact parameter distribution (FF model).
The central (full ) curve in the
 upper band  corresponds to the $BN_\gamma$ model
 with $p_{tmin}=1.3\ GeV, p=0.75$ and GRS densities.
\begin{figure}[htbp] 
   \centering
\includegraphics[scale=0.4]{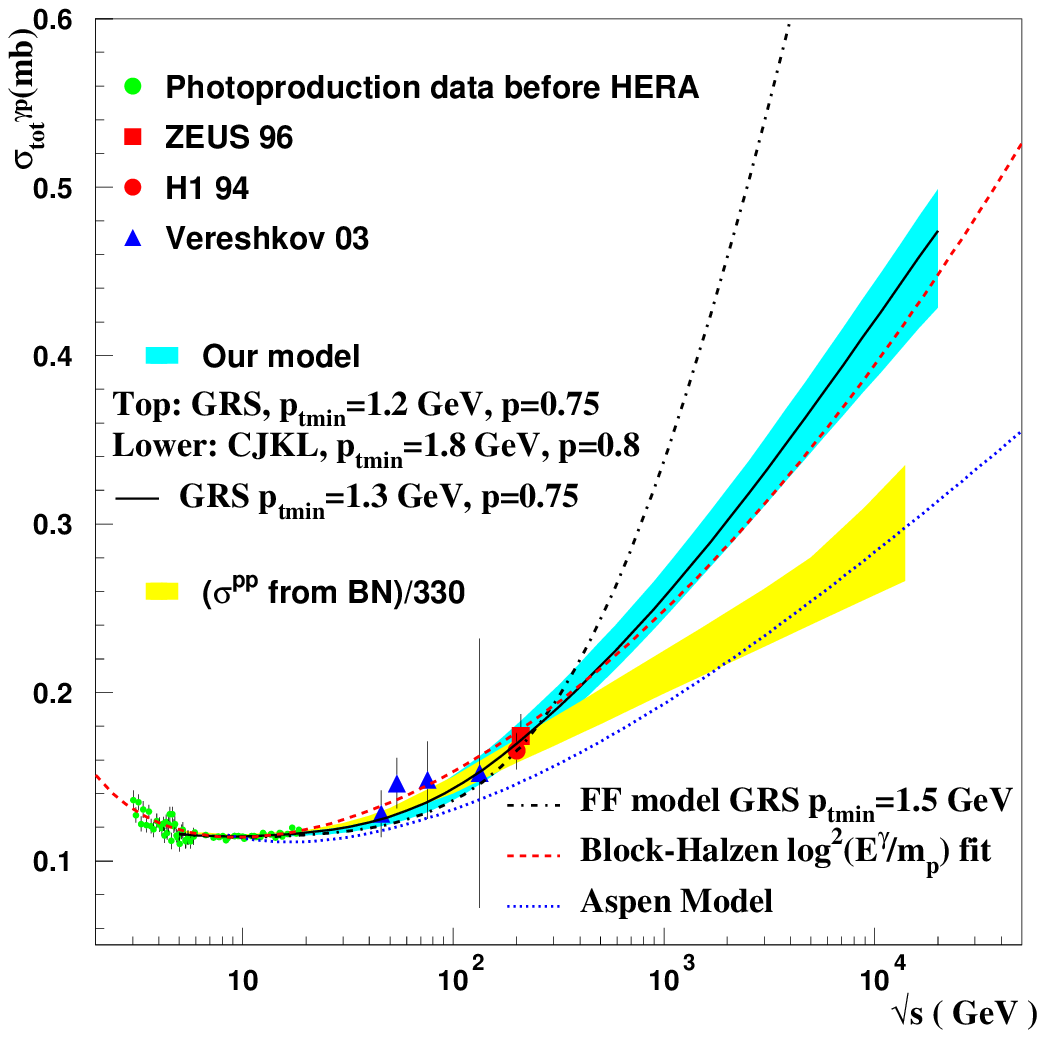}
  \caption{Total $\gamma p$ cross-section with expectation from BN model
using GRS and CJKL densities (upper band) for the photon, compared with the
 model in \cite{aspen},  the fit described in \cite{bhfit},
a factorization model (lower band) and the
eikonal mini-jet model without soft gluons (dot-dashes).}
\vspace{0.2 cm}
   \label{Fig:comparison}
\end{figure}

Fig. \ref{Fig:comparison}  deserves some comment. For the
curves shown in this figure, the parameters have been chosen so as to
reproduce the highest available  accelerator data
(through $p_{tmin}$ and $p$ values for the  BN model,
and through $p_{tmin}$ for the FF model) and the
low energy data, the latter through $P_{had}$ and $\sigma_0$. As the c.m. energy
increases, the model results  show noticeable
differences  between the {\it hadron-like} models,  Aspen
and $BN_F$, and the photon-density model $BN_\gamma$, and much more between
all of them and the eikonal mini-jet (EMM)  Form Factor model. Neglecting the FF model, which we think is
incomplete, we nonetheless have a remarkable difference in the very high
energy range,  $10 \ TeV$ and beyond. Because these predictions may impact
strongly on the photon content of high energy cosmic rays \cite{zas,auger},
this difference does matter.

We notice
that  the  curve, labelled Block-Halzen (BH), from \cite{bhfit}
  lies within   the
 band of the  $BN_\gamma$ model. The BH curve is based on a best fit to
low energy $\gamma p$ data, joined smoothly with a  fit of high energy accelerator
\cite{h1,zeus}  and cosmic ray data
 \cite{vereshkov} of the form
 \begin{equation}
 \sigma_{\gamma p}=c_0+c_1 \log(\nu/m) +c_2\log ^2(\nu/m)+\beta_{{\cal P'}}/\sqrt{\nu/m}
 \end{equation}
 where $\nu$ s the laboratory photon energy.
  There is a noticeable difference between the slope in
  the rising part of the cross-section between the Aspen model and the BH fit,
 as there is between the modelling content between all these descriptions.
In our model the rise  is based on the gluon densities entering the calculation of
the QCD mini-jets cross-sections and on the soft gluon resummation ans\"atz for the impact
parameter distribution. The calculation of these inputs relies
 on realistic PDF distributions and actual,  LO, parton parton cross-section.
 Then, the very high energy (in the TeV region) agreement between the BH  best
 fit based on an analytic expression and our
 results is an independent check of the correct physics content of the BN model.
This fit  confirms
 the inherent interest of our approach based on   QCD mini-jets  and
soft gluon resummation.

For possible use, we  report in table \ref{tablehighsigma}, the numerical values obtained
in our model for the cross-sections shown in Fig. \ref{Fig:comparison}.

\begin{table*}[htb]
\caption{Values (in $m b$) for total cross-section for $\gamma p$
scattering evaluated in the c.m. energy of colliding particles,
corresponding to the bands shown in Fig.~\protect\ref{Fig:comparison}.}
\label{tablehighsigma}
\begin{tabular}{||c||c|c|c|c|c||} \hline \hline
$\sqrt{s} $&EMM with Form &$BN_\gamma$ model &$BN_\gamma$ model
&$BN_{proton}/330$ & $BN_{proton}/330$\\
GeV& Factors,GRS & (upper   curve)& (lower curve)&(upper curve)
&(lower curve)\\
&$p_{tmin}=1.5$ GeV& top band&top band&lower band&lower band\\
\hline \hline
5  &0.116 & 0.116 & 0.116 &0.118  &0.119 \\ \hline
10 &  &  & &0.115  &0.116 \\
11.46  & 0.114 &0.115  & 0.114 & & \\ \hline
48.93 & 0.122 & 0.130    & 0.121  & & \\
50  &  &  &  & 0.131 &0.129  \\ \hline
100 & &  &  &0.15 & 0.143  \\
112.14 & 0.139 &0.155 & 0.140 & & \\ \hline
478.74  & 0.238 &0.228 &0.203 & & \\
500 & & & &0.199 & 0.182 \\ \hline
1000 & & &  &0.221  & 0.199   \\
1097.3  &0.352 &0.279 &0.250 & & \\ \hline
4684.6 & 0.635 &0.384 &0.338 & & \\
5000 & & & &0.280&0.240  \\ \hline
9000 & & & & 0.310 &0.255 \\ \hline
10736.8& 0.829 &0.449 &0.390 & & \\ \hline
14000 && & &0.335 & 0.266  \\ \hline
20000 &0.985 & 0.499 & 0.429 & &\\ \hline \hline
\end{tabular}
\end{table*}

 From a numerical point of view, the curves for the BN  model for protons
and photons  differ because the b-distributions for protons or photons
obtained in our model from  $A_{BN}(q_{max}(s),b)$ differ, and they differ
because the maximum momentum allowed to individual soft gluons is different.
This quantity for the hard part is obtained through the kinematic constraint
averaged over the quark densities and the latter  are of course different for
protons and photons. This is apparent from a comparison of $q_{max}(s)$ for
$pp $ \cite{lastPLB,kazimierz} and $\gamma p$ :  for comparable c.m. energies
$q_{max}(s)$ for $pp$ rises to higher values than the one for $\gamma p$,
resulting in more saturation for $pp$. These differences are due to the quark
densities entering the averaging process defining $q_{max}(s)$: densities are
a phenomenologically extracted quantity and as such it is to be expected that
they reflect the different  structure of the interacting particles, namely the
difference between valence quarks bound in a proton and quark pairs in which
the photon will split and their  respective  evolution.
 \section{Conclusions}
 We have applied to $\gamma p$ scattering an eikonal mini-jet  model with
soft gluon resummation developed for the proton total cross-section. The model
relies on the parton structure of protons and photons and indicates a
different high energy behaviour for $\gamma p$ relative to $pp$ and
${\bar p}p$.  We suggest  that this different behaviour may  be due to
the different parton structure   and high energy evolution properties of
quarks in the proton and quarks in the photon. Furthermore, this result
strengthens our confidence in the BN model as a good approximation to a
QCD description of hadronic interactions in  minimum bias processes.

\section*{Acknowledgments}
We thank Dieter Haidt and Bernd Surrow for pointing out to us the ZEUS BPC95 data. R.G. aknowledges support from the Department of Science and Technology, India, under the J.C. Bose fellowship. G. P.  thanks  the Boston University Theoretical Physics group and the MIT LNS for hospitality
while this work was being written.
This work has been partially supported by MEC (FPA2006-\-05294) and  Junta
de
Andaluc\'\i a (FQM 101 and FQM 437).

\renewcommand{\theequation}{A\arabic{equation}}
\setcounter{equation}{0}
\section*{Appendix A: The mini-jet cross-section}
The QCD jet cross-section
for the process
\begin{equation}
hadron_A+hadron_B\rightarrow X + jet
\end{equation}
is obtained by embedding the parton-parton subprocess cross-section with the given parton densities and integrating over all values of incoming parton momenta and outgoing parton transverse momentum $p_t$, according to the expression
\begin{eqnarray}
\sigma^{AB}_{\rm jet} (s,p_{tmin})=
\int_{p_{tmin}}^{\sqrt{s}/2} \!\! d p_t \int_{4
p_t^2/s}^1 \! \!\! d x_1  \int_{4 p_t^2/(x_1 s)}^1 \! \! \! \! d x_2  \nonumber \\
\times \sum_{i,j,k,l}
f_{i|A}(x_1,p_t^2) f_{j|B}(x_2,p_t^2)
  \frac { d \hat{\sigma}_{ij}^{ kl}(\hat{s})} {d p_t} \ \ \ \ \ \
  \end{eqnarray}
where $A$ and $B$ are the colliding hadrons or photons, in this case $A-proton, B-\gamma$. By construction,  this cross-section  depends on the particular parametrization of the DGLAP \cite{DGLAP} evoluted parton densities, some of which do extend to very low x-values but not too high $p_t^2$ values. This cross-section strongly depends on the lowest $p_t$ value on which one integrates. The term {\it mini-jet} was introduced long ago \cite{jacob,rubbia} to indicate all those low $p_t$ processes  which one can
still expect to be QCD calculable but which are actually not observed as hard jets.  $p_t$ being the scale at which to evaluate $\alpha_s$ in the mini-jet cross-section calculation, one can have  $p_{tmin}\approx 1\div 2\ GeV$.

\renewcommand{\theequation}{B\arabic{equation}}
\setcounter{equation}{0}
\section*{Appendix B: The calculation of $q_{max}(s)$}
Simple kinematics can give the maximum transverse momentum allowed to single gluon emission in a process like
\begin{equation}
parton_1(x_1)+parton_2(x_2)\rightarrow gluon (k)+X(Q)
\end{equation}
namely
\begin{equation}
M(x_1,x_2,Q^2)=
{{\sqrt{\hat s}}
\over{2}}
(1-{{ Q^2}\over{\hat s }})
\end{equation}
with ${\hat s}=s x_1 x_2$.
If  X represents two jets from the outgoing parton-antiparton pair, one can use  $Q^2\approx 4p_t^2$.
The calculation is simplified by introducing  an average over the parton parton cross-section  and integrate over all $x$ values \cite{mario} obtaining
\begin{equation}
q_{\max } (s)   ={{ \sqrt {s}}\over{2}}
\frac{ {\int (dx_1  dx_2)  \int_{z_{min}}^1 \! \! \! \! \! \! dz
\sqrt {x_1 x_2 } (1 - z) D(x_1,x_2)} } {\int (dx_1 dx_2) \int_{z_{min}}^1
 {dz} D(x_1,x_2)}
\end{equation}
where $z_{min}=4p_{tmin}^2/s$, $D$ denotes the usual quark density expression
\begin{equation}
D(x_1,x_2) = {\sum\limits_{i,j}} [f_i (x_1)/x_1] [f_j (x_2)/x_2]
\end{equation}
and we have also assumed that the parton-parton cross-section, appearing at both numerator and denominator, can be evaluated at its maximum value, $p_t=p_{tmin}$, thus dropping out of the calculation.



\begin{thebibliography}{99}
\bibitem{lastPLB}
  A.~Achilli, R.~M.~Godbole, A.~Grau, R.~Hegde, G.~Pancheri and Y.~Srivastava,
  Phys. Lett. B {\bf 659} (2008) 137  [arXiv:0708.3626 (hep-ph)].
\bibitem{martin} For a recent review, see M. Block, Phys.Rept.\textbf{436} (2006) 71-215.
e-Print: hep-ph/0606215, and references therein;
 M.~M.~Block and F.~Halzen,
  Phys. Rev.  D {\bf 73} (2006) 054022  [arXiv:hep-ph/0510238];
\bibitem{REVPHOTSTR}
For reviews see, for example,
M.~Drees and R.~M.~Godbole, J. Phys. G {\bf 21} (1995) 1559
[arXiv:hep-ph/9508221]; M.~Krawczyk, A.~Zembrzuski and M.~Staszel,
  Phys. Rep. {\bf 345} (2001) 265  [arXiv:hep-ph/0011083].
 \bibitem{rohinicorsetti}
 A.~Corsetti, R.~M.~Godbole and G.~Pancheri,
  Phys.\ Lett.\  B {\bf 435} (1998) 441
  [arXiv:hep-ph/9807236].
\bibitem{albert}
R.~M.~Godbole, A.~De Roeck, A.~Grau and G.~Pancheri,
  JHEP {\bf 0306} (2003) 061
  [arXiv:hep-ph/0305071].
\bibitem{lincoll}
 R.~M.~Godbole and G.~Pancheri,
  Eur.\ Phys.\ J.\  C {\bf 19} (2001) 129
  [arXiv:hep-ph/0010104].
\bibitem{gribov} V. N. Gribov, J. Exp. The. Phys. (USSR) vol. {\bf 41} (1961) p. 667.
English translation JETP vol.{\bf 14} (1962) 478; V. N. Gribov and I. Ya.
Pomeranchuk, Phys. Rev. Lett. {\bf  8} (1962) 343.
\bibitem{martinfacto} M. Block and  K.Kang, Int. J. Mod. Phys. {\bf A20} (2005) 27812794.
e-Print: hep-ph/0302146; M.M. Block and A.B. Kaidalov, Phys. Rev
. {\bf D64} (2001) 076002, e-print: hep-ph/0012365.
\bibitem{cudell} 
J.R. Cudell, E. Martynov, G. Soyez, Nucl.Phys. {\bf B682} (2004):391 [arXiv:  hep-ph/0207196];
J.~R.~Cudell and O.~V.~Selyugin,
Phys.Lett. {\bf B662} (2008) 417
[ arXiv:hep-ph/0612046].
\bibitem{bsw} C. Bourrely, J. Soffer and T.T. Wu,
 Mod. Phys. Lett. {\bf A15} (2000), 9-13.  [arXiv: hep-ph/9903438]
 \bibitem{dl} A.~Donnachie and P.~V.~Landshoff,
  Phys. Lett. B {\bf 296} (1992) 227  [arXiv:hep-ph/9209205];
  A.~Donnachie and P.~V.~Landshoff,
  Phys. Lett.  B {\bf 595}(2004) 393  [arXiv:hep-ph/0402081].
 \bibitem{h1}
 H1 Collaboration, S. Aid et al.,  Zeit. Phys.
\ {\bf C69} (1995) 27, hep-ex/9509001.
\bibitem{zeus}
ZEUS collaboration, S. Chekanov et al.,
Nucl. Phys. {\bf  B627} (2002) 3, hep-ex/0202034.
\bibitem{ourlast}  R.~M.~Godbole, A.~Grau, G.~Pancheri and Y.~N.~Srivastava,
Phys. Rev. D {\bf 72} (2005) 076001  [arXiv:hep-ph/0408355].
\bibitem{froissart} M. Froissart, Phys. Rev. {\bf 123} (1961) 1053.\\
 A. Martin, Phys. Rev. {\bf 129} (1963) 1432; Nuovo Cimento {\bf 42} (1966) 930.
 \bibitem{GLM} E. Gotsman, E. Levin and U. Maor, Eur.Phys.J. {\bf C5} (1998) 303,
hep-ph/9708275.
\bibitem{datagg} L3 Collaboration,
M. Acciarri,  et al, {\bf CERN-EP/2001-012}, Phys. Lett. {\bf B519} (2001) 33
, hep-ex/0102025;
OPAL Collaboration.
G. Abbiendi et al., Eur. Phys. J. {\bf C14} (2000) 199 .
\bibitem{PDG}
W.-M. Yao {\em et al.} {\bf Particle Data Group}, J. Phys. G {\bf 33} (2006) 1.
\bibitem{dataproton}
G. Arnison  {\em et al.}, {\bf UA1} Collaboration, Phys. Lett. {\bf
128B} (1983) 336 ; R.~Battiston {\it et al.}  {\bf UA4}
Collaboration, Phys.\ Lett. {\bf B117} (1982) 126;  C.~Augier {\it
et al.}  {\bf UA4/2} Collaboration, Phys.\ Lett. {\bf B344} (1995) 451
; M. Bozzo {\em et al.} {\bf UA4} Collaboration,  Phys. Lett.
{\bf 147B} (1984) 392 ; G.J. Alner {\em et al.} {\bf UA5}
Collaboration, Z.\ Phys. {\bf   C32} (1986) 153 ; N.~Amos {\em
et.~al.}, {\bf E710} Collaboration, Phys. Rev. Lett. {\bf 68} (1992) 2433
; C.~Avila {\em et.~al.}, {\bf E811}
Collaboration, Phys. Lett. {\bf B445} (1999) 419; F.~Abe
{\em et.~al.}, {\bf CDF} Collaboration, Phys. Rev. {\bf D50} (1994) 5550.
\bibitem{collins} J.C.~Collins and G.A.~Ladinsky, Phys. Rev.{\bf D43} (1991) 2847.
\bibitem{PDG08} C. Amsler {\em et al.}, The Review of Particle Physics, Phys. Lett.{\bf B667} (2008) 1.
\bibitem{desy} ECFA/DESY LC Physics Working Group (E. Accomando et al.),
Phys. Rep. {\bf 299} (1998) 1 [arXiv: hep-ph/9705442].
\bibitem{vereshkov08}
Y. Novoseltsev, R. Novoseltseva, G. Vereshkov,  J. Phys. {\bf G36} (2009) 025009,
e-Print: arXiv:0802.0956 [hep-ph].
\bibitem{aspen} M. Block, E. Gregores, F. Halzen and G. Pancheri, Phys.
Rev. {\bf D60} (1999) 054024, e-Print: hep-ph/9809403.
\bibitem{martincosmic}
M.M. Block, F. Halzen, T. Stanev, Phys. Rev.D {\bf 62} (2000) 077501, 
e-Print: hep-ph/0004232.
\bibitem{our99}A.~Grau,
G.~Pancheri and Y.~N.~Srivastava,
Phys. Rev.  D {\bf 60} (1999) 114020 [arXiv:hep-ph/9905228].
\bibitem{corsetti} A.~Corsetti, A.~Grau, G.~Pancheri and Y.~N.~Srivastava,
Phys. Lett. B {\bf 382} (1996) 282 [arXiv:hep-ph/9605314].
  \bibitem{chengwu}
H. Cheng and   T. T. Wu, Phys.Rev. {\bf 186} (1969) 1611.
 \bibitem{bswimpact} 
C. Bourrely, J. Soffer , T. T. Wu,  Phys. Rev. D {\bf 19} (1979) 3249.
\bibitem{jacob} R. Horgan and M. Jacob, Nucl. Phys. {\bf B179}  (1981) 441.
\bibitem{rubbia} G. Pancheri and C. Rubbia, Nucl. Phys. {\bf A} (1984)
418:117C-138C.
\bibitem{YFS} 
D.R. Yennie, S. C. Frautschi and   H. Suura,  Annals Phys. {\bf 13} (1961) 379.
 \bibitem{ddt}Y. I.  Dokshitzer, D.I. Dyakonov and S.I. Troyan, Phys. Lett. {\bf 79B} (1978) 269.
\bibitem{pp} G. Parisi and R. Petronzio, Nucl.Phys. {\bf B154} (1979) 427.
\bibitem{halzenscott} 
F. Halzen, A. D. Martin, D.M. Scott, M.P. Tuite, Z. Phys. {\bf C14} (1982) 351.
\bibitem{altarelli} 
G. Altarelli, R.K. Ellis, M. Greco, G. Martinelli, Nucl.Phys. {\bf B246} (1984) 12.
\bibitem{CGS}
G. Curci, Mario Greco, Y. Srivastava,  Nucl.Phys. {\bf B159} (1979) 451.
\bibitem{ourkt} G. Pancheri-Srivastava and Y.N. Srivastava, Phys.Rev. {\bf D15} (1977) 2915.
\bibitem{dokshitzer} Y.L. Dokshitzer, {\it Perturbative QCD Theory (includes our knowledge of
$\alpha_s$)}, Plenary Talk at ICHEP 1998, Vancouver, hep-ph/9812252
\bibitem{sarcevic} 
Raj Gandhi, Ina Sarcevic,  Phys. Rev. {\bf D44} (1991) 10-14.
\bibitem{halzen} R.S.~Fletcher, T.K. Gaisser and F.~Halzen,
 Phys. Lett. {\bf B298} (1993) 442;  Phys. Rev. {\bf D45} (1992) 377-381, Erratum-ibid.{\bf D45} (1992) 3279.
\bibitem{GRV} M.~Gl\"uck, E.~Reya, and A.~Vogt,  Z. Phys. {\bf C53}(1992) 127;
Z. Phys. {\bf C67} (1995) 433;  Eur. Phys. J. {\bf C 5} (1998) 461.
\bibitem{MRST} A.~D. Martin, R.~G. Roberts, W.~J. Stirling, and R.~S. Thorne,
Phys.  Lett. {\bf B531} (2002) 216.
\bibitem{CTEQ}  H.L. Lai, J. Botts, J.
Huston, J.G. Morfin, J.F. Owens, Jian-wei Qiu, W.K. Tung, H.
Weerts, Phys.Rev. {\bf D51} (1995) 4763.
\bibitem{GRVPHO}M.~Gl\"uck, E.~Reya and A.~Vogt, Phys. Rev.{\bf D 46} (1992)
1973.
\bibitem{GRS} M. Gl\"uck, E. Reya and I. Schienbein, Phys. Rev. {\bf D 60}
(1999) 054019; Erratum, ibid {\bf D 62} (2000) 019902.
\bibitem{CJKL}  F. Cornet, P. Jankowski, M. Krawczyk and A. Lorca,
Phys. Rev. {\bf D 68} (2003) 014010.
\bibitem{photon07} R.M. Godbole, A. Grau, G. Pancheri and Y.N. Srivastava, Nucl. Phys. Proc. Supp. {\bf 184} (2008) 85-90, 
 e-Print: arXiv:0802.3367 [hep-ph].
\bibitem{durand} 
L. Durand, H. Pi, Phys.Rev.{\bf D40} (1989) 1436.
\bibitem{ourepjc}
R. M. Godbole , G. Pancheri, Eur.Phys.J.{\bf C 19} (2001) 129,
e-Print: hep-ph/0010104.
\bibitem{photon03}
{\it Photon total cross-sections},
R.M. Godbole, A. Grau , G. Pancheri, Y.N. Srivastava,  Nucl. Phys. Proc. Suppl.
{\bf 126} (2004) 94-99. Also in {\it Frascati 2003, The structure and
interactions of the photon}, p.94-99
e-Print: hep-ph/0311211.
\bibitem{luna}
E.G.S. Luna,  Phys. Lett. {\bf B641} (2006) 171-176.
e-Print: hep-ph/0608091, and references therein; E.G.S. Luna and A.A. Natale,
Phys. Rev. {\bf D 73} (2006) 074019, hep-ph/0602181.
\bibitem{bartels}
J. Bartels, D. Colferai, S. Gieseke, A. Kyrieleis,  Phys.Rev. {\bf D66} (2002) 094017.
e-Print: hep-ph/0208130; J. Bartels, S. Gieseke and C.F. Qiao, Phys. Rev. {\bf D63} (2001) 056014, Erratum-ibid.{\bf D65} (2002)079902,
e-Print: hep-ph/0009102.
\bibitem{ETP} E. Etim, G. Pancheri and B. Touschek, Nuovo Cimento {\bf 51B} (1967) 276.
\bibitem{ptintrinsic} 
A. Nakamura, G. Pancheri, Y.N. Srivastava,  Z. Phys. {\bf C21} (1984) 243.
\bibitem{mario} M. Greco and P. Chiappetta,
Nucl. Phys. {\bf B221} (1983) 269.
\bibitem{yndurain} 
F.J. Yndurain,
Lectures given at 17th Autumn School: {\it QCD: Perturbative or Nonperturbative?}
(AUTUMN 99), Lisbon, Portugal, 29 Sep - 4 Oct 1999.
Published in {\it Lisbon 1999, QCD: Perturbative or nonperturbative?} p.97-129
e-Print: hep-ph/9910399.
\bibitem{richardson} J. L. Richardson, Phys. Lett. {\bf B82} (1979) 272.
For an application to deep inelastic scattering, see K. Adel, F. Barreiro and F. J. Yndurain, Nucl. Phys.{\bf B495} (1997) 221.
\bibitem{polyakov} A.M.  Polyakov, JETP Lett. {\bf 20} (1974) 194.
\bibitem{dieter} D. Haidt , {\it The transition from $\sigma(\gamma^* p)$ to $\sigma(\gamma p)$},
Prepared for 9th International Workshop on Deep Inelastic Scattering (DIS 2001), Bologna, Italy, 27 Apr - 1 May 2001.  Published in {\it Bologna 2001, Deep inelastic scattering} 287-290, and refs. therein.
\bibitem{bpc}  ZEUS Collaboration, J. Breitweg et al., Phys. Let. {\bf B407} (1997) 432, DESY-97-135, hep-ex/9707025v3;  ZEUS Collaboration, J. Breitweg et al., Phys. Let. {\bf B487} (2000) 53, DESY-00-071, hep-ex/0005018v2.
\bibitem{bernd} B. Surrow, DESY-THESIS-1998-004; A. Bornheim, in the
{\it Proceedings of the LISHEP International School on High Energy Physics,
Brazil, 1998}, hep-ph/9806021.
\bibitem{bpcdata} ZEUS Collaboration, J. Breitweg et al., EPJC {\bf 7} (1999) 609, DESY-98-121, hep-ex/9809005v1.
\bibitem{vereshkov}
G.M. Vereshkov, O.D. Lalakulich, Yu.F. Novoseltsev, R.V. Novoseltseva,
Phys.Atom.Nucl.{\bf 66} (2003) 565-574, Yad.Fiz.{\bf 66} (2003) 591-600.
\bibitem{zas} 
M. Ave, J.A. Hinton, R.A. Vazquez, A.A. Watson, E. Zas, Phys. Rev. {\bf D67}
(2003) 043005.
e-Print: astro-ph/0208228;
M. Ave, J.A. Hinton, R.A. Vazquez , A.A. Watson, E. Zas, Phys. Rev. {\bf D65}
(2002) 063007.
e-Print: astro-ph/0110613.
\bibitem{auger} 
Pierre Auger Collaboration (J. Abraham et al.). FERMILAB-PUB-07-736-A, Dec 2007. 28pp.
Astropart.Phys. {\bf 29} (2008)  e-Print: arXiv:0712.1147 [astro-ph].
\bibitem{bhfit} M. Block and F. Halzen,
 Phys.Rev. {\bf D70} (2004) 091901. e-Print: hep-ph/0405174; ibidem
Phys. Rev. {\bf D72} (2005) 036006, Erratum-ibid.{\bf D72} (2005)039902.
e-Print: hep-ph/0506031.
\bibitem{kazimierz}
G. Pancheri, R.M. Godbole, A. Grau, Y.N. Srivastava,
Acta Phys. Polon. B {\bf 38} (2007) 2979 [arXiv:hep-ph/0703174].
\bibitem{DGLAP} Y. L. Dokshitzer, Sov. Phys. JETP {\bf 46} (1977) 641
[Zh. Eksp. Teor. Fiz.{\bf 73}
(1977) 1216]; V. N. Gribov and L. N. Lipatov, Yad. Fiz. {\bf 15} (1972) 781
[Sov.
J. Nucl. Phys. {\bf 15} (1972) 438]; G. Altarelli and G. Parisi, Nucl. Phys.
{\bf B126} (1977) 298.
\end{thebibliography}
%
%
%


\end{document}